\documentclass{aastex631}

\usepackage{amsmath}
\usepackage{empheq}
\usepackage{rotating}

\shortauthors{A. Sicilia et al.}
\shorttitle{(Super)Massive BHMF}

\begin{document}

\title{The Black Hole Mass Function Across Cosmic Times\\
II. Heavy Seeds and (Super)Massive Black Holes}

\author[0000-0002-4515-3540]{Alex Sicilia}\affiliation{SISSA, Via Bonomea 265, 34136 Trieste, Italy}

\author[0000-0002-4882-1735]{Andrea Lapi}
\affiliation{SISSA, Via Bonomea 265, 34136 Trieste, Italy}\affiliation{IFPU - Institute for fundamental physics of the Universe, Via Beirut 2, 34014 Trieste, Italy}\affiliation{INFN-Sezione di Trieste, via Valerio 2, 34127 Trieste,  Italy}\affiliation{IRA-INAF, Via Gobetti 101, 40129 Bologna, Italy}

\author[0000-0003-3127-922X]{Lumen Boco}\affiliation{SISSA, Via Bonomea 265, 34136 Trieste, Italy}\affiliation{IFPU - Institute for fundamental physics of the Universe, Via Beirut 2, 34014 Trieste, Italy}

\author[0000-0001-8973-5051]{Francesco Shankar}\affiliation{Department of Physics and Astronomy, University of Southampton, Highfield SO17 1BJ, UK}

\author[0000-0002-5896-6313]{David M. Alexander}\affiliation{Department of Physics, Durham University, South Road, Durham, DH1 3LE, UK}

\author[0000-0001-7232-5152]{Viola Allevato}\affiliation{INAF-Osservatorio di Astrofisica e Scienza dello Spazio di Bologna, 40129 Bologna, Italy}\affiliation{Scuola Normale Superiore, Piazza dei Cavalieri 7, 56126 Pisa, Italy}

\author[0000-0002-8956-6654]{Carolin Villforth} \affiliation{Dept. of Physics, Univ. of Bath, Claverton Down, BA27AY, Bath, UK}

\author[0000-0002-0375-8330]{Marcella Massardi}\affiliation{IRA-INAF, Italian ARC, Via Gobetti 101, I-40129 Bologna, Italy}
\affiliation{SISSA, Via Bonomea 265, 34136 Trieste, Italy}

\author[0000-0003-0930-6930]{Mario Spera}\affiliation{SISSA, Via Bonomea 265, 34136 Trieste, Italy}\affiliation{IFPU - Institute for fundamental physics of the Universe, Via Beirut 2, 34014 Trieste, Italy}\affiliation{INFN-Sezione di Trieste, via Valerio 2, 34127 Trieste,  Italy}

\author[0000-0002-7922-8440]{Alessandro Bressan}\affiliation{SISSA, Via Bonomea 265, 34136 Trieste, Italy}\affiliation{IFPU - Institute for fundamental physics of the Universe, Via Beirut 2, 34014 Trieste, Italy}

\author[0000-0003-1186-8430]{Luigi Danese}\affiliation{SISSA, Via Bonomea 265, 34136 Trieste, Italy}\affiliation{IFPU - Institute for fundamental physics of the Universe, Via Beirut 2, 34014 Trieste, Italy}

\begin{abstract}
This is the second paper in a series aimed at modeling the black hole (BH) mass function, from the stellar to the (super)massive regime. In the present work we focus on (super)massive BHs and provide an ab-initio computation of their mass function across cosmic times. We consider two main mechanisms to grow the central BH, that are expected to cooperate in the high-redshift star-forming progenitors of local massive galaxies. The first is the gaseous dynamical friction process, that can cause the migration toward the nuclear regions of stellar-mass BHs originated during the intense bursts of star formation in the gas-rich host progenitor galaxy, and the buildup of a central heavy BH seed $M_\bullet\sim 10^{3-5}\, M_\odot$ within short timescales $\lesssim$ some $10^7$ yr. The second mechanism is the standard Eddington-type gas disk accretion onto the heavy BH seed, through which the central BH can become (super)massive $M_\bullet\sim 10^{6-10}\, M_\odot$ within the typical star-formation duration $\lesssim 1$ Gyr of the host. We validate our semi-empirical approach by reproducing the observed redshift-dependent bolometric AGN luminosity functions and Eddington ratio distributions, and the relationship between the star-formation and the bolometric luminosity of the accreting central BH. We then derive the relic (super)massive BH mass function at different redshifts via a generalized continuity equation approach, and compare it with present observational estimates. Finally, we reconstruct the overall BH mass function from the stellar to the (super)massive regime, over more than ten orders of magnitudes in BH mass.
\end{abstract}


\setcounter{footnote}{0}

\section{Introduction}\label{sec|intro}

The formation of (super)massive black holes (BHs) with masses $M_\bullet\sim 10^{6-10}\, M_\odot$ and their role in galaxy evolution constitute crucial yet longstanding problems in modern astrophysics and cosmology. These monsters are thought to have grown mainly by gaseous accretion onto a disk surrounding the BH (e.g., Lynden-Bell 1969; Shakura \& Sunyaev 1973) that energizes the spectacular broadband emission of active galactic nuclei (AGNs), and leaves a BH relic ubiquitously found at the center of massive galaxies in the local Universe (e.g., Kormendy \& Ho 2013; also textbooks by Mo et al. 2010, Cimatti et al. 2020). This paradigm has recently received an astonishing confirmation by the EHT collaboration (2019, 2022) via the imaging of the BH shadow caused by gravitational light bending and photon capture at the event horizon of M87 and Sgr A$^\star$.

Accreting supermassive BHs can have a profound impact on the evolution of the host galaxies (see review by Alexander \& Hickox 2012), as testified by the observed tight relationships between the relic BH masses and the physical properties of the hosts, most noticeably the stellar mass or velocity dispersion of the bulge component (e.g., Magorrian 1998; Ferrarese \& Merritt 2000; Gebhardt et al. 2000; Tremaine et al. 2002; Kormendy \& Ho 2013; McConnell \& Ma 2013; Reines \& Volonteri 2015; Sahu et al. 2019; Shankar et al. 2016, 2020a; Zhu et al. 2021). These suggest that (apart from short-time stochastic fluctuations) the BH and the bulge stellar mass must have co-evolved over comparable timescales, possibly determined by the energy feedback from the BH itself on the gas/dust content of the host (see Tinsley 1980; Silk \& Rees 1998; Fabian 1999; King 2005; Lapi et al. 2006, 2014, 2018; for a review, see King \& Pounds 2015). In fact, targeted X-ray observations in the high-redshift star-forming progenitors of local massive galaxies have started to reveal the early growth of a dust-enshrouded (super)massive BH in their nuclear regions (e.g., Mullaney et al. 2012; Page et al. 2012; Delvecchio et al. 2015; Rodighiero et al. 2015, 2019; Fiore et al. 2017; Stanley et al. 2015, 2017; Massardi et al. 2018; Combes et al. 2019; D'Amato et al. 2020), before it attains a high enough mass and power to manifest as a bright AGN and to eventually reduce/quench star formation and partly evacuate gas and dust from the host (e.g., Granato et al. 2001, 2004; Lapi et al. 2014, 2018). Another, albeit more indirect, indication of coevolution for the bulk of the BH and the host stellar mass comes from the similarity between the activity timescales of central BH to the transition timescale of (green valley) galaxies from the blue cloud to the red sequence (see Wang et al. 2017; Lin et al. 2021, 2022; this is true apart from rejuvenations at late cosmic times, see Martin-Navarro et al. 2021).

However, two recent pieces of evidence may suggest that standard disk accretion is not the only process at work in growing a BH to the (super)massive regime. The first is the discovery of an increasing number of active BHs with masses $M_\bullet\gtrsim 10^9\, M_\odot$ at very high redshifts $z\gtrsim 7$ (e.g., Mortlock et al. 2011; Wu et al. 2015; Venemans et al. 2017; 2018; Reed et al. 2019; Banados et al. 2018, 2021; Wang et al. 2019, 2021), when the age of the Universe was shorter than $0.8$ Gyr. The second is the robust measurements of extreme BH masses $M_\bullet\gtrsim 10^{9-10}\, M_\odot$ at the center of early-type galaxies with stellar mass $M_\star\gtrsim 10^{11}\, M_\odot$ (e.g., McConnell et al. 2011; Ferre-Mateu et al. 2015; Thomas et al. 2016; Mehrgan et al. 2019; Dullo et al. 2021), that have formed most of their old stellar component during a star-formation episode lasting some $10^8$ yr at $z\gtrsim 1$, as demonstrated by astro-archeological measurements of their stellar ages and $\alpha$-enhanced metal content (e.g., Thomas et al. 2005, 2010; Gallazzi et al. 2006, 2014; Johansson et al. 2012; Maiolino \& Mannucci 2019; Morishita et al. 2019; Saracco et al. 2020). These observations concur to raise the issue of how billion solar mass BHs may have grown in less than a Gyr. In fact, this is somewhat challenging if standard disk accretion starts from a light seed $\sim 10^2\, M_\odot$ of stellar origin and proceeds with the typical Eddington ratios $\lambda\lesssim 1$ as estimated out to $z\sim 6$ in active BHs (see Vestergaard \& Osmer 2009; Nobuta et al. 2012; Kelly \& Shen 2013; Dai et al. 2014; Kim \& Im 2019; Duras et al. 2020; Ananna et al. 2022), that would require an overall time $\gtrsim 0.8/\lambda$ Gyr to attain $\sim 10^9\, M_\odot$. Solutions may invoke mechanisms able to rapidly produce heavy BH seeds $10^{3-5}\, M_\odot$, so reducing the time required to attain the final masses by standard disk accretion (see Natarajan 2104, Mayer \& Bonoli 2019 and Volonteri et al. 2021 for exhaustive reviews). Viable possibilities comprise: direct collapse of gas clouds within a (proto)galaxy, possibly induced by galaxy mergers or enhanced matter inflow along cosmic filaments (e.g., Lodato \& Natarajan 2007; Mayer et al. 2010, 2015; Di Matteo et al. 2012, 2017); merging of stars inside globular or nuclear star clusters (e.g., Portegies Zwart et al. 2004; Devecchi et al. 2012; Latif \& Ferrara 2016); migration of stellar BHs towards the nuclear galaxy regions via dynamical friction against the dense gas-rich environment in strongly star-forming progenitors of local massive galaxies (e.g., Boco et al. 2020, 2021).
\footnote{When referring to the dynamical friction mechanism, the term 'seeds' is used in a broader sense with respect to the classic meaning in the literature. A seed is usually referred to as the first compact object on which subsequent disk accretion occurs, eventually leading to the formation of a supermassive BH. The heavy seeds formed with the dynamical friction mechanism are by-products of multiple mergers of already-existing stellar-mass BHs (that in turn could be referred as light seeds) forming across a wide redshift range.}

Such a complex picture for the overall (super)massive BH growth may in principle be probed via one of the most fundamental quantities for demographic studies of the BH population, namely the BH mass function, that expresses the number density of BHs per comoving volume and unit BH mass as a function of redshift. For (super)massive BHs, where most of the mass is accumulated through gas disk accretion, this is usually estimated (but still subject to systematics) from the AGN luminosity functions via Soltan (1982)-type or continuity equation arguments (e.g., Small \& Blandford 1992; Haehnelt et al. 1998; Salucci et al. 1999; Yu \& Tremaine 2002; Yu \& Lu 2004, 2008; Merloni \& Heinz 2008; Cao 2010; Kelly \& Merloni 2012; Aversa et al. 2015; Shankar et al. 2004, 2009, 2013), or from local galaxy mass/luminosity/velocity dispersion functions and scaling relation among these properties and the BH mass (e.g., Vika et al. 2009; Li et al. 2011; Mutlu-Pakdil et al. 2016; Shankar et al. 2016, 2020a).

In a future perspective, a precise assessment of the relic BH mass function is also important to work out detailed predictions for the gravitational wave emission expected from mergers of (super)massive BHs, that will constitute the primary targets of the upcoming Laser Interferometer Space Antenna mission (e.g., Sesana et al. 2016; Ricarte \& Natarajan 2018; for a review, Barausse \& Lapi 2021 and references therein) and of ongoing and future Pulsar-Timing Array experiments (e.g., Antoniadis et al. 2022). Thus a theoretical grasp on the (super)massive BH mass function across cosmic times is of crucial importance.

This is the second paper in a series aimed at modeling the BH mass function, from the stellar to the intermediate and (super)massive regime. In Sicilia et al. (2022; hereafter paper I) we have focused on the stellar mass BH relic mass function, while in the present work we provide an ab-initio computation of
the redshift-dependent mass function for (super)massive BHs.  We consider two mechanisms to grow the central BH, that likely cooperate in the high-redshift star-forming progenitors of local massive galaxies. The first one is the gaseous dynamical friction introduced by Boco et al. (2020), which can cause the migration of stellar-mass BHs originated during the intense bursts of star formation in the gas-rich central regions of the host progenitor galaxy, and the buildup of heavy BH seeds $\lesssim 10^5\, M_\odot$ within short timescales $\lesssim$ some $10^7$ yr. The second mechanism is the standard Eddington-type gas disk accretion onto the
heavy seed, through which the central BH can become (super)massive within the typical
star-formation timescales $\lesssim 1$ Gyr of the host galaxy. 

Our approach is semi-empirical, requires minimal modeling and a few educated assumptions, and is original in at least three respects: (i) we start from the galaxy SFR functions and derive BH-related statistics by jointly modeling the evolution of the central BH mass and the stellar mass of the host; (ii)
we explicitly compute (and do not assume a priori) the heavy seed mass function by exploiting the distribution of stellar mass BHs originated from star-formation (Sicilia et al. 2022) and their migration rates due to dynamical friction (Boco et al. 2020); (iii) we determine the detailed shape of the BH growth curve during disk accretion (in particular, we set the Eddington ratio) by requiring that the final BH and host stellar mass satisfy a Magorrian-like relationship, and that the star-formation timescale of the galaxy host is set by the main-sequence relation. We validate our approach by reproducing the observed redshift-dependent bolometric AGN luminosity functions and Eddington ratio distributions, and the relationship between the star-formation of the host galaxy and the bolometric luminosity of the accreting central BH. We then derive the relic (super)massive BH mass function at different redshifts via a generalized continuity equation approach, and compare it with present observational estimates. At the same time, we provide a robust theoretical basis for a physically-motivated heavy seed distribution at high redshifts. Finally, we put together the results from paper I and the present work to reconstruct the overall BH mass function from the stellar to the intermediate to the (super)massive regime, over more than ten orders of magnitudes in BH mass. 

The plan of the paper is straightforward: in Section \ref{sec|basics} we describe our semi-empirical framework, in Section \ref{sec|results} we present and discuss our results, and in Section \ref{sec|summary} we summarize our main findings and outline future perspectives. In the Appendix we recall the basics of the gaseous dynamical friction mechanism (see Appendix \ref{sec|app_dynfric}) and of the continuity equation technique (see Appendix \ref{sec|app_conteq}) exploited in the computations of the main text.
Throughout this work, we adopt the standard flat $\Lambda$CDM cosmology (Planck Collaboration 2020) with rounded parameter values: matter density $\Omega_{\rm M}=0.3$, dark energy density $\Omega_{\Lambda}=0.7$, baryon density $\Omega_{\rm b}=0.05$, Hubble constant $H_0=100\, h$ km\ s$^{-1}$ Mpc$^{-1}$ with $h=0.7$, and mass variance $\sigma_8=0.8$ on a scale of $8\, h^{-1}$ Mpc. A Kroupa (2001) initial mass function (IMF) in the star mass range $m_\star\sim 0.1-150\, M_\odot$ is adopted. 

\section{Theoretical background}\label{sec|basics}

We aim to derive the redshift-dependent (super)massive relic BH mass function ${\rm d}N/{\rm d}\log M_\bullet\, {\rm d}V$, i.e., the number density of massive BHs per unit comoving volume $V$ and BH mass $M_\bullet$. We rely on two main mechanisms to grow the central BH mass, which are likely to cooperate in the gas-rich star-forming progenitors of local massive galaxies (hosting massive relic BHs): gas disk accretion and stellar BH migration via gaseous dynamical friction. Both processes will require the joint modeling of the stellar and BH mass growth history in a galaxy of given SFR and redshift. 

\subsection{Stellar mass growth}

As to the stellar mass growth, we assume a simple two-stage star formation history
\begin{equation}\label{eq|SFR}
\dot M_{\star}(\tau) = (1-\mathcal{R})\,\psi\, \Theta_{\rm H}[\tau\leq \tau_b] \end{equation}
where $\psi$ is the SFR, $\mathcal{R}\approx 0.45$ is the IMF-dependent gas fraction restituted to the ISM during stelllar evolution (with the quoted value applying for a Kroupa IMF), and $\Theta_{\rm H}[\cdot]$ is the Heaviside step function. Basically, this reflects a constant SFR, that is then abruptly quenched by the radiative/kinetic power associated to SN explosions/stellar winds and/or the central BH activity at around the age $\tau_{\rm b}$. This temporal evolution renders to a good approximation the behavior expected from state-of-the-art in-situ galaxy evolution models (e.g., Pantoni et al. 2019; Lapi et al. 2020), and is also indicated by SED-modeling studies of high-redshift dusty star-forming galaxies (e.g., Papovich et al. 2011; Smit et al. 2012; Moustakas et al. 2013; Steinhardt et al. 2014; Cassar\'a et al. 2016; Citro et al. 2016; Gonzalez Delgado et al. 2017; Carnall et al. 2019;  Williams et al. 2021; Pantoni et al. 2021), and by the observed fraction of IR-detected host galaxies in X-ray (e.g., Mullaney et al. 2012; Page et al. 2012; Rosario et al. 2012; Azadi et al. 2015; Stanley et al. 2015;  Carraro et al. 2020) and in IR or optically selected AGNs (e.g., Mor et al. 2012; Wang et al. 2013; Willott et al. 2015; Stanley et al. 2017; Dai et al. 2018; Bianchini et al. 2019; Nguyen et al. 2020; Wang et al. 2021). 

Correspondingly, the stellar mass increases as
\begin{equation}\label{eq|Mstar}
M_{\star}(\tau) = \left\{
\begin{aligned}
&(1-\mathcal{R})\,\psi\, \tau   & ,\; \tau\leq \tau_{\rm b} \\
\\
& M_\star(\tau_{\rm b}) & ,\; \tau> \tau_{\rm b}
\end{aligned}
\right.
\end{equation}
and hereafter we will indicate for convenience $M_{\star,\rm relic}\equiv M_\star(\tau_{\rm b})=(1-\mathcal{R})\,\psi\, \tau_{\rm b}$. Note, however, that this is the stellar mass just before the quenching at $\tau_{\rm b}$, not the relic stellar mass at $z\approx 0$; in fact, at late cosmic times this may be further increased by dry mergers, especially in very massive galaxies (e.g., Rodriguez-Gomez et al. 2016; Buitrago
et al. 2017; Lapi et al. 2018).

We can estimate the value of the star-formation duration $\tau_{\rm b}$ by requiring that, just before the quenching, the SFR $\psi$ and the stellar mass $M_\star(\tau_{\rm b})=M_{\star,\rm relic}$ satisfy the redshift-dependent main sequence relationship $\psi_{\rm MS}(M_\star,z)$ (see Daddi et al. 2007; Rodighiero et al. 2011, 2015; Sargent et al. 2012; Speagle et al. 2014; Whitaker et al. 2014; Schreiber et al. 2015; Caputi et al. 2017; Bisigello et al. 2018; Boogaard et al. 2018; Leja et al. 2022; Rinaldi et al. 2022; Popesso et al. 2022); in other words, the condition
\begin{equation}\label{eq|taub}
\psi_{\rm MS}(M_{\star, \rm relic},z)=\psi
\end{equation}
sets the timescale $\tau_{\rm b}(\psi,z)$ for any galaxy with SFR $\psi$ and redshift $z$. We adopt as our reference the main-sequence determination by Speagle et al. (2014)
\begin{equation}\label{eq|MS_sp14}
\log \frac{\psi_{\rm MS}(M_\star,z)}{M_\odot\, {\rm yr}^{-1}} \approx (-6.51+0.11\, t_z)+(0.84-0.026\,t_z)\, \log \frac{M_\star}{M_\odot}\;.
\end{equation}
where $t_z$ is the age of the Universe at redshift $z$ in units of Gyr. We will show in Sect. \ref{sec|comparison} the effect of adopting a different main sequence prescription.

\subsection{BH growth due to dynamical friction}

Boco et al. (2020, 2021) have pointed out that the central BH can grow, especially in the early stages, by a continuous rain of stellar mass BHs that are funnelled toward the nuclear region via dynamical friction against the gas-rich background of high-redshift star-forming galaxies. The related growth rate is computed following Boco et al. (2020; see also Appendix A), to which we refer the interested reader for details. For consistency, in the present work we initialize the computation basing on the stellar BH mass function and light seed distribution derived in paper I, along the following lines.

First of all, we extract from the stellar and binary evolutionary code \texttt{SEVN} (see Spera et al. 2019) the so-called stellar term, i.e. the number of BHs originated per unit of stellar mass formed $M_{\rm SFR}$ and BH mass $m_\bullet$:
\begin{equation}\label{eq|stellarterm}
\cfrac{{\rm d} N_\bullet}{{\rm d}M_{\rm SFR}{\rm d}\log m_{\bullet}}(m_\bullet|Z)=\cfrac{{\rm d}N_{\star\rightarrow \bullet}}{{\rm d}M_{\rm SFR}{\rm d}\log m_{\bullet}}+\cfrac{{\rm d}N_{\star\star\rightarrow \bullet}}{{\rm d}M_{\rm SFR}{\rm d}\log m_{\bullet}}+\sum_{i=1,2}\cfrac{{\rm d}N_{\star\star\rightarrow \bullet\bullet}}{{\rm d}M_{\rm SFR}{\rm d}\log m_{\bullet,i}}\;.
\end{equation}
This includes three different contributions from isolated stars evolving into BHs ($\star\rightarrow \bullet$), from stars that are originally in binary systems but end up as an isolated BH because one of the companions has been ejected or destroyed or cannibalized ($\star\star\rightarrow \bullet$), and from stars in binary systems that evolve into binary BHs ($\star\star\rightarrow \bullet\bullet$). All these terms are strongly dependent on metallicity $Z$, which affects the efficiency of the various processes involved in stellar and binary evolution, like mass loss rates, mass transfers, core-collapse physics, etc. (see paper I for details).

We then derive the birthrate of stellar BHs with mass $m_\bullet$ at time $\tau$ in an individual galaxy with SFR $\psi$ at redshift $z$ from the expression
\begin{equation}\label{eq|Rbirth}
\frac{{\rm d}\dot N_{\rm birth}}{{\rm d}\log m_\bullet}(m_\bullet,\tau|\psi,z) = \psi\,\int{\rm d}\log Z\,\cfrac{{\rm d}N_\bullet}{{\rm d}M_{\rm SFR}{\rm d}\log m_{\bullet}}(m_\bullet|Z)\, \frac{{\rm d}p}{{\rm d}\log Z}[Z|Z_{\rm FMR}(\psi,M_\star(\tau))]\;.
\end{equation}
The integrand is the product of the stellar term from the previous Eq. (\ref{eq|stellarterm}) and of the metallicity distribution ${\rm d}p/{\rm d}\log Z$. For the latter we adopt a lognormal shape centered around the fundamental metallicity relation $\log Z_{\rm FMR}(\psi,M_\star)$ by Mannucci et al. (2011; for a review, see Maiolino \& Mannucci 2019), with a dispersion of $\Delta\log Z_{\rm FMR}\approx 0.15$ dex, and $M_\star(\tau)$ given by Eq. (\ref{eq|Mstar}).

We then compute the migration rate per unit stellar BH mass due to gaseous dynamical friction at time $\tau$ inside a galaxy with SFR $\psi$ at redshift $z$
\begin{equation}\label{eq|dynfricrate}
\frac{{\rm d}\dot N_{\rm DF}}{{\rm d}\log m_\bullet}(m_\bullet,\tau|\psi,z)=\int{\rm d} r\frac{{\rm d} p}{{\rm d} r}(r)\int{\rm d} v_\theta\frac{{\rm d} p}{{\rm d} v_\theta}(v_\theta|r)\,
\int{\rm d} v_r\frac{{\rm d} p}{{\rm d} v_r}(v_r|r)\,\frac{{\rm d}\dot N_{\rm birth}}{{\rm d}\log m_\bullet}(m_\bullet,\tau-\tau_{\rm DF}|\psi,z)~;
\end{equation}
here ${\rm d} p/{\rm d} r$ and ${\rm d} p/{\rm d} v_{r,\theta}$ are the probability distributions of initial radii and velocities and $\tau_{\rm DF}(m_\bullet, r, v_r,v_\theta)$ is the dynamical friction timescale against the gaseous background for a compact remnant of mass $m_\bullet$. All these quantities are recalled in Appendix A and detailed in Boco et al. (2020); in the latter paper the reader can find an account of how the dynamical friction timescale depends on such quantities and on the parameters ruling the gas distribution.

Finally, the growth rate of the central BH due to the dynamical friction mechanism is just
\begin{equation}\label{eq|BHdynfricrate}
\dot{M}_{\bullet, \rm DF}(\tau|\psi,z)=\Theta_{\rm H}[\tau\leq \tau_{\rm b}]\,\int{\rm d}\log m_\bullet\, m_\bullet\, \frac{{\rm d}\dot N_{\rm DF}}{{\rm d}\log m_\bullet}(m_\bullet,\tau|\psi,z)~,
\end{equation}
where the step function $\theta_{\rm H}(\cdot)$ specifies that the mechanism is no longer active after $\tau_{\rm b}$ since the gaseous medium is expected to have been at least partly evacuated from the nuclear regions due to feedback processes.

\subsection{BH growth due to gas accretion}

In parallel, the central BH can grow due to standard gas disk accretion. We adopt a BH accretion rate curve with shape (e.g., Yu \& Lu 2004, 2008; Shen 2009; Li 2012; Lapi et al. 2014; Aversa et al. 2015)
\begin{equation}\label{eq|BHaccrate}
\dot M_{\bullet,\rm acc}(\tau) = \left\{
\begin{aligned}
&\frac{M_\bullet(\tau)}{\tau_{\rm ef}}   & , \; \tau\leq \tau_{\rm b} \\
\\
&\dot M_{\bullet, \rm acc}(\tau_{\rm b})\, e^{-(\tau-\tau_{\rm b})/\tau_{\rm d}} & , \; \tau_{\rm b}<\tau\leq \tau_{\rm b}+\zeta\,\tau_{\rm d}\\
\\
& 0 & , \; \tau> \tau_{\rm b}+\zeta\,\tau_{\rm d}
\end{aligned}
\right.
\end{equation}
This describes a growth due to disk accretion in two stages, separated at the galaxy age $\tau_{\rm b}$ where star formation is quenched. The rationale behind the above expression is the following: at early epochs (ages $\tau\lesssim \tau_{\rm b}$) when there is plenty of material to accrete onto the BH in the nuclear galaxy regions, a demand-limited, Eddington-type BH accretion rate over a characteristic $e-$folding timescale $\tau_{\rm ef}$ is assumed. At late times (ages $\tau\gtrsim \tau_{\rm b}$) the BH mass and radiative/kinetic power may be so large as to quench the star formation and partly evacuate gas from the host; however, if residual gas mass
is still present in the central regions, it can be accreted in a supply-driven fashion, thus originating the exponentially declining part of the accretion curve with a characteristic timescale $\tau_{\rm d}$. The IR-detected fraction of X-ray selected AGNs (seee Mullaney et al. 2012; Page et al. 2012; Rosario et al. 2012; Azadi et al. 2015; Stanley et al. 2015; Carraro et al. 2020) suggests $\tau_{\rm d}\approx 2\, \tau_{\rm ef}$, as shown by Lapi et al. (2014) and adopted by Aversa et al. (2015) and Mancuso et al. (2016b, 2017). Eventually, we consider the accretion to stop for ages $\tau\gtrsim\tau_{\rm b}+\zeta\, \tau_{\rm d}$ with $\zeta\approx 3$ (our results are anyway weakly affected by the value of this latter quantity); this is reasonable since at that point the accretion rate becomes so small with $\dot M_{\bullet,\rm acc}\, \tau_{\rm ef}/M_{\bullet}\lesssim 10^{-2}$ as to enter in an ADAF (i.e., advection-dominated accretion flow) regime, where the mass growth can be safely neglected with respect to that accumulated during the slim/thin disk accretion. We will show in Sect. \ref{sec|comparison} the effect of adopting a different, scale-free declining portion (e.g., Shen 2009) of the BH growth curve, that also avoids the inclusion of the quantity $\zeta$.

Provided that $L=\epsilon\, \dot M_\bullet\, c^2$ is the accretion luminosity and $\lambda\equiv L/L_{\rm Edd}$ the (luminous) Eddington ratio in terms of the Eddington luminosity\footnote{The Eddington luminosity $L_{\rm Edd}\equiv 4\pi\, G\, \mu_e\, m_p\, c\, M_\bullet/\sigma_T\approx 1.3\times 10^{38}\, (M_\bullet/M_\odot)$ erg s$^{-1}$ is the limiting value for which the continuum radiation force emitted by the accretion disk balances gravity in isotropic conditions; in the above definition $G$ is the gravitational constant,  $\mu_e$ is the mean molecular weight per electron, $m_p$ is the proton mass, $c$ is the speed of light, and $\sigma_T$ is the Thomson cross section.} $L_{\rm Edd}$, the $e-$folding time of the early growth reads
\begin{equation}\label{eq|tauef}
\tau_{\rm ef} = \frac{\epsilon}{(1-\epsilon)\, \lambda}\, t_{\rm Edd}\; ,
\end{equation}
where $t_{\rm Edd}=M_\bullet\, c^2/L_{\rm Edd}\approx 0.45$ Gyr is the Eddington timescale and $\epsilon$ is the radiative efficiency. As to the latter, it is worth considering that in the early stages the demand-limited accretion may be prone to the development of a slim accretion disk (e.g., Abramowicz et al. 1988), while at late-times the supply-limited accretion tends to originate a classic thin-disk accretion (e.g., Shakura \& Sunyaev 1973).
To describe both conditions we express the radiative efficiency via the prescription by Aversa et al. (2015) valid for both thin and slim disks (see also Mineshige et al. 2000; Watarai et al. 2001; Li 2012; Madau et al. 2014):
\begin{equation}\label{eq|radeff_av15}
\epsilon\approx \epsilon_{\rm thin}\, \frac{\lambda/2}{e^{\lambda/2}-1}\; ;
\end{equation}
here $\epsilon_{\rm thin}$ is the efficiency during the thin-disk phase, which may range from $\approx 0.057$ for a nonrotating BH to $\approx 0.32$ for a maximally rotating Kerr BH (see Thorne 1974). We will adopt
$\epsilon_{\rm thin}\approx 0.15$ as our fiducial value (see Davis \& Laor 2011; Raimundo et al. 2012; Trakhtenbrot et al. 2017; Shankar et al. 2020b), but will show in Sect. \ref{sec|comparison} the effect of adopting a larger efficiency $\epsilon_{\rm thin}\approx 0.3$. Note that in principle $\epsilon_{\rm thin}$ may even depend on the galactic age since in the early stages the accretion is likely chaotic and so the spin of the BH should stay rather small, while in the late stages a coherent accretion is expected  to set in and the spin can rapidly increase to maximal values (see Lapi et al. 2014); however, we neglect such spin/efficiency evolution in the present framework. 

\subsection{Overall BH growth}

The overall growth of the central BH mass due to both dynamical friction and gaseous accretion writes
\begin{equation}\label{eq|BHrate}
\dot M_\bullet(\tau)=\dot{M}_{\bullet, \rm DF}(\tau)+\dot{M}_{\bullet,\rm acc}[M_\bullet(\tau)]\;.
\end{equation}
Given Eqs. (\ref{eq|BHdynfricrate}) and (\ref{eq|BHaccrate}), the previous equation can be formally integrated to yield the overall central BH mass growth
\begin{equation}\label{eq|BHmass}
M_{\bullet}(\tau|\psi,z) = \left\{
\begin{aligned}
& \int_{0}^\tau{\rm d}\tau'\, e^{(\tau-\tau')/\tau_{\rm ef}}\,\dot M_{\rm DF}(\tau') & \tau\leq \tau_{\rm b}\\
\\
& M_\bullet(\tau_{\rm b})\, \left[1+\frac{\tau_{\rm d}}{\tau_{\rm ef}}\,(1-e^{-(\tau-\tau_{\rm b})/\tau_{\rm d}})\right] & \tau_{\rm b}<\tau\leq \tau_{\rm b}+\zeta\,\tau_{\rm d}\\
\\
& M_\bullet(\tau_{\rm b})\, \left[1+\frac{\tau_{\rm d}}{\tau_{\rm ef}}\,(1-e^{-\zeta})\right] & \tau> \tau_{\rm b}+\zeta\, \tau_{\rm d}
\end{aligned}
\right.
\end{equation}
where the value on the last line corresponds to the final, relic BH mass $M_{\bullet, \rm relic}(\psi,z) = M_\bullet(\tau_{\rm b})\, [1+(\tau_{\rm d}/\tau_{\rm ef})\times (1-e^{-\zeta})]$.

The dynamical friction process dominates in the initial growth stage for $\tau\ll \tau_{\rm ef}$; we will show that it provides, by inducing the migration of stellar BHs originated from star formation, heavy seeds of order $10^{3-5}\, M_\odot$ within some $10^7$ yr, before standard Eddington-type accretion takes over as the dominant mechanism for BH growth. Remarkably, our modelling above, at variance with other approaches in the literature, does not require assumptions regarding the seed BH from which to start gas accretion: the light seeds are provided by star formation and stellar evolution, and the heavy seeds by the gaseous dynamical friction mechanism, in a consistent way.

As a consequence, for any galaxy with SFR $\psi$ at redshift $z$, the evolution of the BH mass is completely specified by assigning the Eddington factor $\lambda$ of the early growth stage, which determines the radiative efficiency $\epsilon$ via Eq. (\ref{eq|radeff_av15}) and hence the $e-$folding timescale $\tau_{\rm ef}$. We empirically determine the Eddington ratio (see Shankar et al. 2020b) by requiring that the relic BH and stellar mass just after the quenching at $\tau_{\rm b}$ satisfy a Magorrian-like relation $M_{\bullet,{\rm Mag}}(M_\star,z)$, with a possible redshift dependence. In other words, from the condition
\begin{equation}\label{eq|lambda}
M_{\bullet, \rm relic}(\psi,z)=M_{\bullet,{\rm Mag}}[M_{\star,\rm relic}(\psi,z),z]
\end{equation}
one can determine $\lambda(\psi,z)$ for any galaxy with SFR $\psi$ and redshift $z$. We rely on the debiased determination of the Magorrian relationship by Shankar et al. (2016, 2020a) 
\begin{equation}\label{eq|MBHMbulge}
\log \frac{M_{\bullet,{\rm Mag}}}{M_\odot}(M_\star,z) \approx 7.574+1.946\,\log\frac{M_\star}{10^{11}\, M_\odot}-0.306\, \log^2\frac{M_\star}{10^{11}\, M_\odot}-0.011\, \log^3\frac{M_\star}{10^{11}\, M_\odot}+\eta\,\log(1+z)
\end{equation}
holding in the range $M_\star\sim 10^{10-12}\, M_\odot$, with an intrinsic scatter $\sigma_{\log M_{\bullet,{\rm Mag}}}\approx 0.32-0.1\, \log (M_\star/10^{12}\, M_\odot)$ dex. The parameter $\eta$ in the above equation allows for a possible evolution with redshift $M_{\bullet,{\rm Mag}}(M_\star,z)\propto (1+z)^\eta$; this is considerably debated in the literature (e.g., Merloni et al. 2010; Schulze \& Wisotzki 2014; Ding et al. 2020; Suh et al. 2020; Li et al. 2021; Habouzit et al. 2022), but the latest studies suggest a mild evolution with $\eta\approx 0.2$, that we take as our fiducial value (our results are anyway weakly affected by this choice). We will show in Sect. \ref{sec|comparison} the effect of adopting a different Magorrian relationship. 

We stress that at least two low-redshift processes, that can in principle affect the BH mass function, have not been considered in our framework: (i) relic supermassive BHs can be reactivated by accretion of gas funnelled toward the central regions by galaxy mergers or internal disk instabilities (e.g., Di Matteo et al. 2005; Capelo et al. 2015), that can trigger spectacular radio-mode activity in terms of relativistic jets; (ii) relic supermassive BHs can coalesce following, with some delay, a galaxy merger. In fact, the impact of these processes on the supermassive BH mass function is still somewhat debated: an important role of galaxy mergers in reproducing the massive end of the mass function has been claimed in semi-analytic models (e.g., Marulli et al. 2008; Bonoli et al. 2009), while other semi-empirical and numerical approaches have instead pointed out a much more limited relevance of mergers on the BH mass function (e.g., Aversa et al. 2015; Steinborn et al. 2018; McAlpine et al. 2020). The detailed treatment of galaxy and BH mergers is beyond the main scope of the present paper, and is deferred to future work.

\subsection{BH growth rate function}

Toward a statistical description, we start from the SFR function ${\rm d}N/{\rm d}\log\psi\, {\rm d}V$, i.e., the number density of galaxies with given SFR $\psi$ per unit comoving cosmological volume $V$ at redshift $z$. For this we adopt the determination by Boco et al. (2021, their Fig. 1; for an analytic Schechter fit see Eq. 2 and Table 1 in Mancuso et al. 2016a) derived from an educated combination of the dust-corrected UV (e.g., Oesch et al. 2018; Bouwens et al. 2021), IR (e.g., Gruppioni et al. 2020; Zavala et al. 2021), and radio (e.g. Novak 2017; Ocran 2020) luminosity functions, appropriately converted into SFR (see Kennicutt \& Evans 2012) using our assumed Kroupa (2001) IMF.

We first compute the central BH growth rate function
\begin{equation}\label{eq|GRF}
\frac{{\rm d} N}{{\rm d}\log \dot M_\bullet\, {\rm d}V}(\dot M_\bullet,z) = \int{\rm d}\log\psi \cfrac{{\rm d}N}{{\rm d}\log\psi\,{\rm d}V}(\psi,z)\, \cfrac{1}{\tau_{\rm b}}\, \sum_i \cfrac{{\rm d}\tau_i}{{\rm d}\log \dot M_\bullet}(\dot M_\bullet|\psi,z)\; ,
\end{equation}
where $\dot M_\bullet(\tau|\psi,z)$ is provided by Eq. (\ref{eq|BHmass}) and 
${\rm d}\tau/{\rm d}\log \dot M_\bullet$ is the related time spent by the BH in a logarithmic bin of given growth rate; the summation allows for multiple solutions $\tau_i$ of the equation $\dot M_\bullet(\tau|\psi,z)=\dot M_\bullet$, that are typically two for the growth curve assumed in this work. Note that the SFR dependence in $\dot M_\bullet(\tau|\psi,z)$ is twofold: on the one hand it is related to the growth rate of heavy seeds by migration of stellar-mass BHs, whose birthrate ultimately depends on star formation; on the other hand, such a dependence is encoded in the Eddington ratio $\lambda(\psi,z)$ derived after Eq. (\ref{eq|lambda}) and in the radiative efficiency given by Eq. (\ref{eq|radeff_av15}). To allow for some scatter induced by the Magorrian-like relationship, one can write
\begin{equation}\label{eq|GRFscatter}
\begin{aligned}
\sum_i \cfrac{{\rm d}\tau_i}{{\rm d}\log \dot M_\bullet} & = \int_0^{\tau_{\rm b}+\zeta\,\tau_{\rm d}}{\rm d}\tau\; 
\delta_{\rm D}\left[\log \dot M_\bullet-\log \dot M_\bullet(\tau|\psi,z)\right]\simeq \\
\\
&\simeq \int_0^{\tau_{\rm b}+\zeta\,\tau_{\rm d}}{\rm d}\tau\;\cfrac{{\rm sech}^2\left\{[\log\dot M_\bullet-\log \dot M_\bullet(\tau|\psi,z)]/2\,\tilde\sigma_{\log\dot M_\bullet}\right\}}{4\, \tilde\sigma_{\log\dot M_\bullet}}\\
\end{aligned}
\end{equation}
the first equality follows trivially from the properties of the Dirac $\delta_{\rm D}[\cdot]$ function, while in the second we have substituted a log-logistic distribution with dispersion $\tilde \sigma_{\log\dot M_\bullet}\simeq (\sqrt{3}/\pi)\, \sigma_{\log\dot M_\bullet}$ in terms of the standard log-normal dispersion $\sigma_{\log\dot M_\bullet}$.
The reason for using a log-logistic distribution in place of the standard log-normal one is that having heavier tails it tends to maintain intrinsic power-law distributions at the high-mass end, as indicated by the data relating to the AGN luminosity functions and BH mass function (see discussion by Ren \& Trenti 2021). Agreement with the latter statistics requires to adopt $\sigma_{\log\dot M_\bullet}\approx 0.3-0.4$ dex, in line with the scatter of the Magorrian.

\subsection{AGN luminosity functions, Eddington ratios and mean SFRs}

The broadband emission of AGNs is energized by the gas accretion onto the (super)massive BHs; thus a relevant statistics to validate our semi-empirical approach is the redshift-dependent bolometric AGNs luminosity function. This may be computed analogously to Eq. (\ref{eq|GRF}) as
\begin{equation}\label{eq|AGNLF}
\frac{{\rm d} N}{{\rm d}\log L_{\rm AGN}\, {\rm d}V}(L_{\rm AGN},z) = \int{\rm d}\log\psi \cfrac{{\rm d}N}{{\rm d}\log\psi\,{\rm d}V}(\psi,z)\, \cfrac{1}{\tau_{\rm b}}\, \sum_i \cfrac{{\rm d}\tau_i}{{\rm d}\log L_{\rm AGN}}(L_{\rm AGN}|\psi,z)\; ,
\end{equation}
where the times $\tau_i$ are now determined from the condition $L_{\rm AGN}=\epsilon\, \dot M_{\bullet,\rm acc}\, c^2/(1-\epsilon)$, with the gas accretion curve $\dot M_{\bullet,\rm acc}(\tau|\psi,z)$ specified by Eq. (\ref{eq|BHaccrate}). Notice that the Eddington ratio $\lambda$ and the radiative efficiency $\epsilon$ here are not free parameters but are self-consistenly computed, for any SFR $\psi$ and redshift $z$, by Eqs. (\ref{eq|lambda}) and (\ref{eq|radeff_av15}). We will compare our results with the bolometric luminosity function determination by Shen et al. (2020), reconstructed from a large compilation of rest-frame B-band/UV (e.g., Hopkins et al. 2007; Giallongo et al. 2012; Manti et al. 2017; Kulkarni et al. 2018), soft/hard X-ray (e.g., Fiore et al. 2012; Ueda et al. 2014; Aird et al. 2015a, 2015b; Miyaji et al. 2015), and IR data (e.g., Assef et al. 2011; Lacy et al. 2015) collected in the past decades (see Shen et al. 2020 for details concerning bolometric and obscuration corrections).

Notice that the integrand in Eq. (\ref{eq|AGNLF}) constitutes the number density of galaxies ${\rm d}^2N/{\rm d}\log\psi\,{\rm d}\log L_{\rm AGN}\, {\rm d}V$ per comoving volume in bins of SFR and AGN luminosity. Thus it may be exploited to build up the so-called coevolution plane SFR vs. $L_{\rm AGN}$, and the mean relationship between these two quantities. Finally, the previous expressions can also be adapted to derive the Eddington-ratio distribution by simply substituting $L_{\rm AGN}$ with $\lambda=L_{\rm AGN}/L_{\rm Edd}$. 

\subsection{Relic BH mass function}

To derive the relic (super)massive BH mass function ${\rm d}N/{\rm d}\log M_\bullet\, {\rm d}V$ we exploit a generalized version of the continuity equation (see Yu \& Lu 2004, 2008; Aversa et al. 2015), whose derivation is recalled in Appendix B; this is basically a technique to relate the BH growth functions (or AGN luminosity functions) to the BH mass functions. The outcome reads
\begin{equation}\label{eq|BHMF}
\cfrac{{\rm d}N}{{\rm d}\log M_\bullet\,{\rm d}V}(M_\bullet,z) =  -\int_z^{\infty}{{\rm d}z'}\, \left|\cfrac{{\rm d}t_{z'}}{{\rm d}z'}\right|\; \left.\cfrac{\partial_{\log \dot M_\bullet} \cfrac{{\rm d}N}{{\rm d}\log \dot M_\bullet\,{\rm d}V}(\dot M_\bullet,z')}{\sum_i \cfrac{{\rm d}\tau_i}{{\rm d}\log \dot M_\bullet}\times\partial_{\log \dot M_\bullet}\log M_\bullet} \right|_{\; \dot M_\bullet=\frac{M_\bullet}{\tau_{\rm ef}+\tau_{\rm d}\,(1-e^{-\zeta})}}\; .
\end{equation}
Here the quantity ${\rm d} N/{\rm d}\log \dot M_\bullet\, {\rm d}V$ is the growth rate function from Eq. (\ref{eq|GRF}), all the integrand is computed at the maximum accretion rate for a given relic BH mass $\dot M_\bullet = M_\bullet/[\tau_{\rm ef}+\tau_{\rm d}\,(1-e^{-\zeta})]$, and the various quantities implicitly entering there (e.g., $\lambda$, $\tau_{\rm ef}$, $\tau_{\rm d}$) must be referred to a relic BH mass $M_\bullet$ and redshift $z$.
We stress that in our framework, at variance with many previous approaches based on continuity equation, the input AGN luminosity functions are not just taken from observations, but are derived from the galaxy statistics via Eq. (\ref{eq|AGNLF}). The related relic (super)massive BH mass density can be computed as
\begin{equation}\label{eq|BHdensity}
\rho_\bullet(z) = \int{\rm d}\log M_\bullet\, M_\bullet\, \cfrac{{\rm d}N}{{\rm d}\log M_\bullet\,{\rm d}V}(M_\bullet,z)\; ,
\end{equation}
where typically the integral is taken over BH masses $M_\bullet\gtrsim 10^{6}\, M_\odot$.

\begin{sidewaysfigure}
\centering
\includegraphics[width=\textwidth]{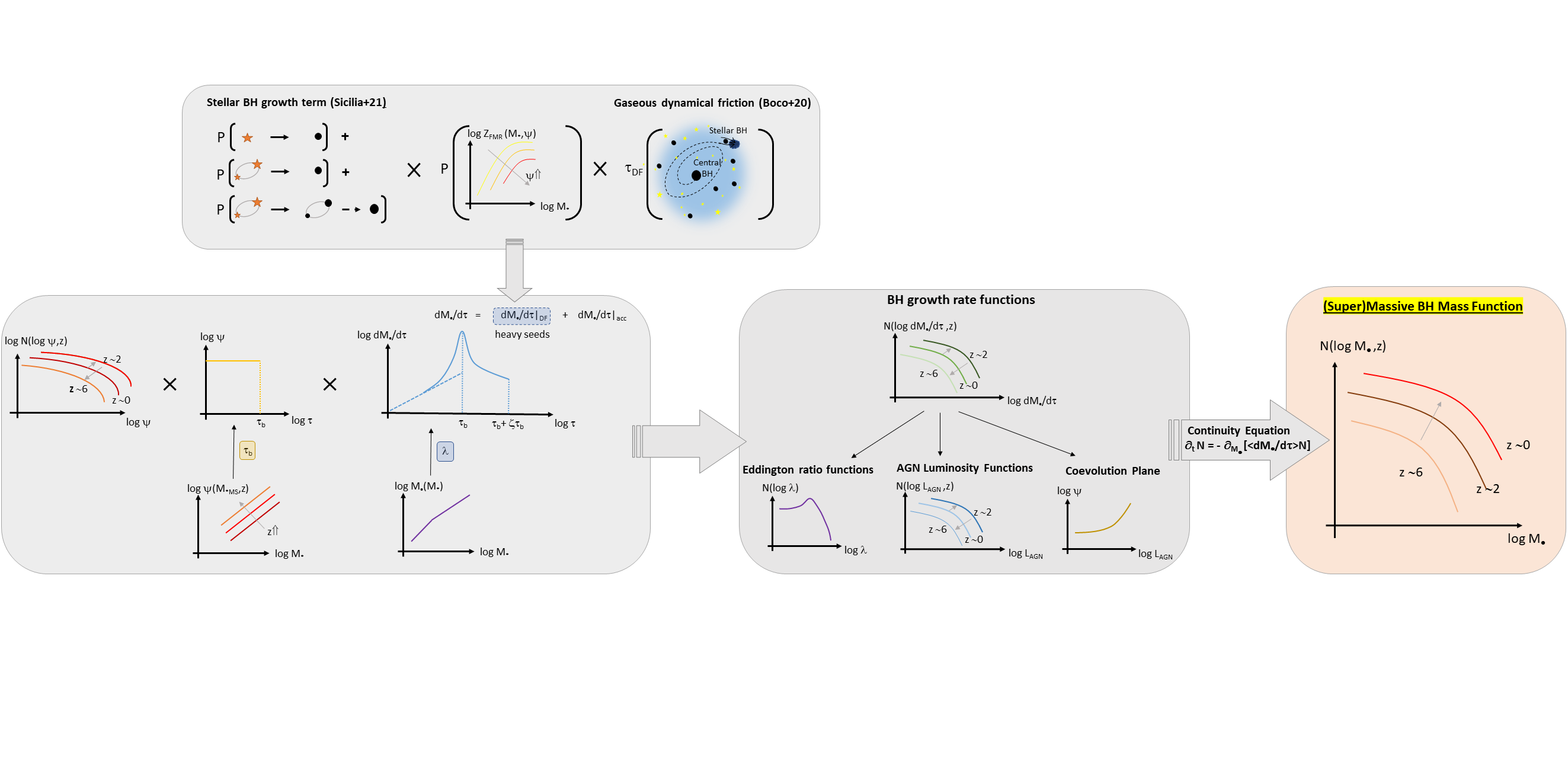}
\caption{Schematics showing the main steps to compute the (super)massive relic BH mass function. The starting point is the stellar term from paper I, representing the number of BHs originated per unit of star formed mass, and includes contributions from the evolution of isolated or binary stars into isolated or binary BHs (light seeds; see Eq. \ref{eq|stellarterm}). This is coupled to the metallicity distribution (extracted from the fundamental metallicity relation) and with the timescale for gaseous dynamical friction by Boco et al. (2020; see Eq. \ref{eq|tauDF}) to derive the growth rate of the central BH by migration of stellar remnants (see Eqs. \ref{eq|dynfricrate} and \ref{eq|BHdynfricrate}). In parallel, galaxy statistics provided by the SFR functions are coupled with model growth curves of the stellar and BH mass (see Eqs. \ref{eq|Mstar} and \ref{eq|BHmass}); the latter includes the growth by dynamical friction migration and by gaseous Eddington-type accretion. Crucial parameters of these growth curves, like the star formation duration and the Eddington factor, are derived by requiring consistency with the main sequence of star-forming galaxies (see Eqs. \ref{eq|taub} and \ref{eq|MS_sp14}) and with the local Magorrian relationship (see Eqs. \ref{eq|lambda} and \ref{eq|MBHMbulge}). The main outcome of this procedure are BH growth rate functions (see Eq. \ref{eq|GRF}), and byproducts are Eddington ratio functions, AGN luminosity functions (Eq. \ref{eq|AGNLF}) and the coevolution plane SFR vs. $L_{\rm AGN}$. Finally, a generalized continuity equation approach allows to convert the growth rate functions into the (super)massive BH mass function (Eq. \ref{eq|BHMF}).}\label{fig|Schematic}
\end{sidewaysfigure}

Fig. \ref{fig|Schematic} summarizes in an illustrative way all the steps followed to compute the (super)massive BH mass function and described in this Section.

\section{Results and Discussion}\label{sec|results}

In this Section we will show results of our empirical model concerning the growth of the central BH mass, AGN luminosity functions and Eddington ratio distribution, relationship of the AGN luminosity with the host SFR, and BH mass function. We will highlight the role played by the gaseous dynamical friction process in providing a physical mechanism to originate heavy seeds, so allowing the growth of the central BH to the supermassive regime at moderate Eddington ratios within the typical star-formation timescale of the host. We will also discuss the dependence of our basic results on various assumptions.

\subsection{Basic results}

\begin{figure}
\centering
\includegraphics[width=\textwidth]{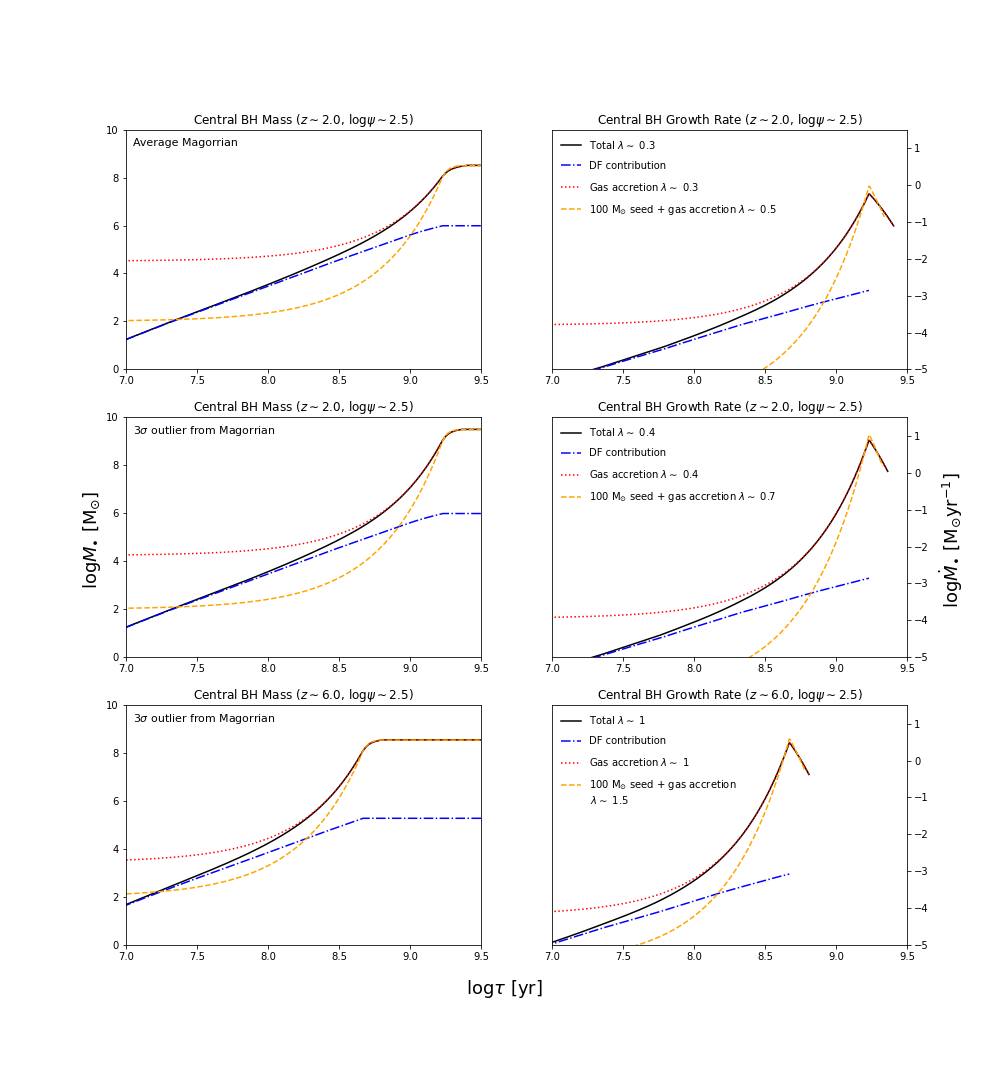}
\caption{Time evolution of the central BH mass (left panels) and BH growth rate (right panels) in a star-forming galaxy with SFR $\psi\sim 300\, M_\odot$ yr$^{-1}$ 
at $z\approx 2$ (top and middle rows) and $z\approx 6$ (bottom row); in the top row the final BH mass is on the average Magorrian relationship, in the middle and bottom rows it is a $3\sigma$ upper outlier of the Magorrian relationship. The overall growth of the central (super)massive BH is illustrated by black solid line (and the corresponding  Eddington ratio $\lambda$ is indicated in the first entry of the legend) while the contribution from migration of stellar BHs via gaseous dynamical friction is shown by the blue dot-dashed line. The red dotted line represents the evolution of a central BH growing by pure disk accretion (i.e., without dynamical friction) with the same final mass and with the same $\lambda$ as the solid line, implying that the initial seed must be $\gtrsim 10^4\, M_\odot$. Finally, the orange dashed line shows the evolution of a central BH growing by pure disk accretion (i.e., without dynamical friction) with the same final mass as the solid line from a stellar mass seed $\approx 100\, M_\odot$, implying an appreciably higher Eddington ratios (indicated in the last entry of the legend).}\label{fig|timevo}
\end{figure}

To start with, in the top row of Fig. \ref{fig|timevo} we illustrate the time evolution of the central BH mass (left panels) and BH growth rate (right panels) in a prototypical star-forming galaxy with SFR $\psi\sim 300\, M_\odot$ yr$^{-1}$ at reference redshifts $z\approx 2$ (top and middle rows) and $z\approx 6$ (bottom row). In the top row the final BH mass is assumed to satisfy the average Magorrian relationship, while in the middle and bottom rows it is taken as a $3\sigma$ upper outlier with respect to the Magorrian; these latter instances are representative of extremely massive BHs, that are possibly sampled because of observational biases (especially at high redshifts). The overall growth is illustrated as black solid lines, and the corresponding Eddington ratio is reported in the first entry of the legend, while the contribution from migration of stellar BHs via gaseous dynamical friction is shown by the blue dot-dashed lines. It is seen that in all these cases the evolution of the total BH mass at small galactic ages is dominated by the growth due to migration of stellar BHs via gaseous dynamical friction; such a process can effectively build up a heavy central BH seed of mass $M_\bullet\sim 10^{3-5}$ within $\lesssim 10^8$ yr. Thereafter Eddington-type gas disk accretion takes over and can grow the central BH to the (super)massive regime $M_\bullet\gtrsim 10^{8-9}\, M_\odot$. Remarkably, the overall effect of the early growth by dynamical friction is twofold. First, it allows the central BH to attain the final mass within a rather short timescale of some $10^8$ yr; this can contribute to alleviate, or even to solve, the high-redshift quasar problem, i.e. the buildup of billion-solar-mass BHs in quasar hosts at $z\gtrsim 6$, when the age of the universe $\lesssim 1$ Gyr constitutes a demanding constraint.
Second, such a growth can be obtained with reasonable values of the Eddington ratios $\lambda\sim 0.3$, that are in sound agreement with the observational determinations (see below); even in the extreme instance of an upper $3\sigma$ outlier of the Magorrian at $z\approx 6$ (bottom panels), the growth can be achieved with sub-Eddington conditions $\lambda\lesssim 1$. 

In addition, in Fig. \ref{fig|timevo} we also illustrate what happens in the absence of the dynamical friction process, hence enforcing a BH growth by pure disk accretion. In particular, the orange dashed lines depict the evolution of a central BH with the same final mass as the solid lines but starting from a stellar mass seed $\approx 10^2\, M_\odot$; such a case is seen to imply an appreciably higher Eddington ratio (reported in the last entry of the legend). In other words, growing the BH from light seeds of stellar origin to the supermassive regime would require a time $\gtrsim 0.8/\lambda$ Gyr. Thus especially at high redshifts and/or for upper outliers of the Magorrian relationship (i.e., BHs with billion solar masses), the growth of the central BH should proceed at appreciably high values of $\lambda$, and possibly in super-Eddington conditions (as in the bottom panels). Though this instance can be partially justified theoretically (e.g., Li 2012; Madau et al. 2014) and there are hints of a few cases at $z\gtrsim 6$ (e.g., Fujimoto et al. 2022), it struggles somewhat against the bulk of present observational estimates at $z\lesssim 6$ (see references below and Fig. \ref{fig|erdf}). On the other hand, the red dotted lines refer to the evolution of a central BH growing to the same final mass and with the same Eddington ratio $\lambda$ as the solid lines; such a case is seen to imply that the initial seeds must be $\gtrsim 10^4\, M_\odot$. Therefore a specific mechanism, alternative to dynamical friction, must be in any case envisaged to obtain such massive seeds (e.g., Volonteri et al. 2021; see also Sect. \ref{sec|intro}).

\begin{figure}
\centering
\includegraphics[width=\textwidth]{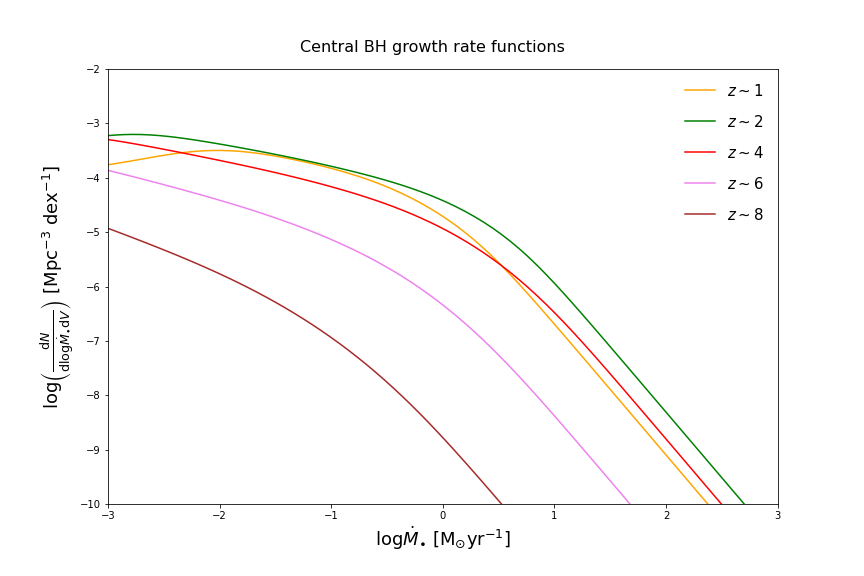}
\caption{The BH growth rate function (see Eq. \ref{eq|GRF}) at different redshifts $z\approx 1$ (orange), $2$ (green), $4$ (red), $6$ (magenta), and $8$ (brown).}\label{fig|BHMF_grf}
\end{figure}

In Fig. \ref{fig|BHMF_grf} we illustrate the growth rate function of the central BH at different redshifts $z\sim 1-8$ (color-coded). As it can be seen from Eq. (\ref{eq|GRF}), its shape as a function of $\dot M_\bullet$ is determined by a combination of galaxy statistics (i.e., the SFR functions) and the time spent by the central BH in a given bin of growth rate. The latter is in turn determined by the shape of the BH growth rate as a function of galactic age plotted in Fig. \ref{fig|timevo}: at early times the BH rate grows almost linearly due to dynamical friction, at intermediate times it raises almost exponentially over the timescale $\tau_{\rm ef}$ due to disk accretion, and at late times it diminishes exponentially over the timescale $\tau_{\rm d}$. The redshift evolution mirrors that of galaxy statistics, with the knee of the function first increasing toward larger $\dot M_\bullet$ out to $z\approx 2$ and then receding at higher redshifts. The turnover of the function at $z\lesssim 1$ at low accretion rates reflects the progressive inefficiency of the dynamical friction process (in turn mirroring the decreased efficiency of star-formation and stellar mass BH generation) toward late cosmic times.

\begin{figure}
\centering
\includegraphics[width=\textwidth]{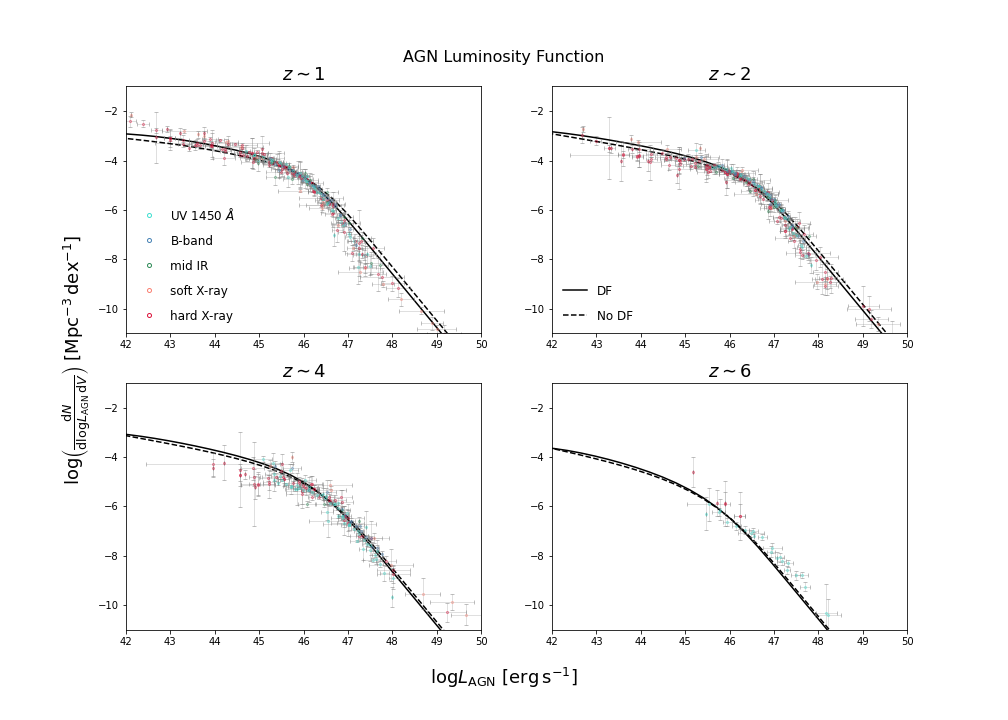}
\caption{The bolometric AGN luminosity functions (see Eq. \ref{eq|AGNLF}) at different redshifts $z\approx 1$ (top left panel), $2$ (top right), $4$ (bottom left) and $6$ (bottom right). Solid lines illustrate the result of our approach, which are compared with the data compilation by Shen et al. (2020; circles) from selections in the UV $1450$ \.{A} (cyan), $B-$band (blue), mid-IR (green), soft X-rays (orange), and hard X-rays (red). For comparison, dashed lines show the results when the gaseous dynamical friction mechanism is switched off (see comment in main text).}\label{fig|agnlf}
\end{figure}

\begin{figure}
\centering
\includegraphics[width=\textwidth]{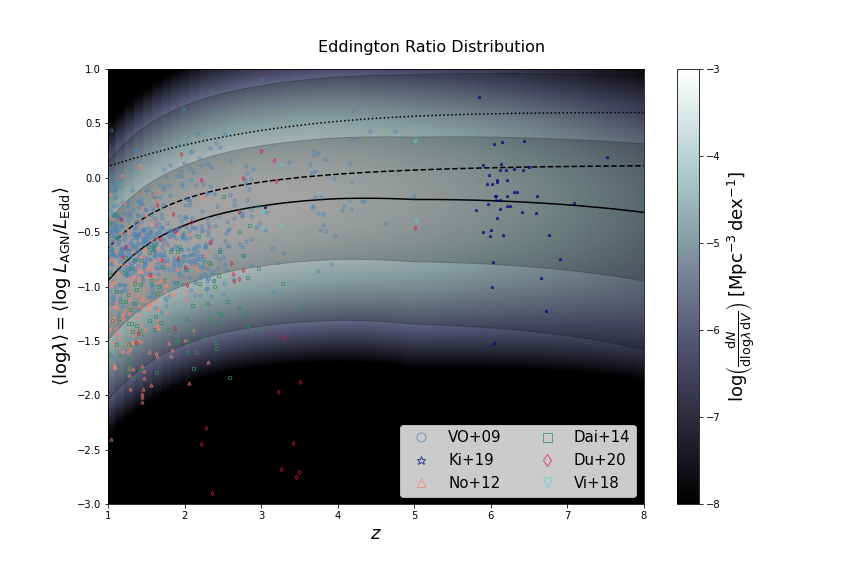}
\caption{The Eddington ratio distribution and average Eddington ratio as a function of redshift $z$. The intensity of the black and white background illustrates the Eddington ratio distribution, while the black solid line is the average relationship expected from our approach (dark and light grey shades represents the $1\sigma$ and $2\sigma$ dispersion). In addition, dashed line is the average Eddington ratio when the gaseous dynamical friction mechanism is switched off, and the dotted line is the average Eddington ratio adopted on an empirical basis by Aversa et al. (2015). Data are from Duras et al. (2020; red diamonds), Kim \& Im (2019; navy stars), Vignali et al. (2018; cyan inverted triangles), Dai et al. 2014 (green squares), Nobuta et al. (2012; orange triangles), and Vestergaard \& Osmer (2009; blue circles).}\label{fig|erdf}
\end{figure}

In Fig. \ref{fig|agnlf} we show the bolometric AGN luminosity functions at different redshifts $z\approx 1$, $2$, $4$ and $6$, computed from Eq. (\ref{eq|AGNLF}). The results from our approach are compared with the  observational estimates collected by Shen et al. (2020; see full list of references therein) from selections in the rest-frame B-band/UV (e.g., Hopkins et al. 2007; Giallongo et al. 2012; Manti et al. 2017; Kulkarni et al. 2018), soft/hard X-ray (e.g., Fiore et al. 2012; Ueda et al. 2014; Aird et al. 2015a, 2015b; Miyaji et al. 2015), and IR (e.g., Assef et al. 2011; Lacy et al. 2015), converted using appropriate bolometric corrections (see Table 1 and Section 3 in Shen et al. 2020). The agreement is pretty good, both in terms of shape and redshift evolution. It is worth mentioning that the number density for AGNs with bright luminosities (especially toward high redshifts) may be overestimated in the data due to the uncertainties in the bolometric corrections. Note that we do not attempt a comparison with the observed AGN luminosity functions at $z\lesssim 1$ since our framework does not include BH reactivations from late-time mergers and disk-instabilities (see Sect. \ref{sec|basics}); the latter are known to be a fundamental ingredient in determining the low-$z$ AGN luminosity functions, especially at the faint end, though reproducing these observables has demonstrated to be a highly non-trivial task even for detailed models incorporating the aforementioned processes (see Griffin et al. 2019; Izquierdo-Villalba et al. 2020).

In Fig. \ref{fig|erdf} we illustrate the Eddington ratio distribution ${\rm d}N/{\rm d}V\, {\rm d}\log\lambda$ and the average Eddington ratio $\langle\log \lambda\rangle$ with its dispersion as a function of redshift. This is compared with observational estimates from different samples (see Duras et al. 2020; Kim \& Im 2019; Vignali et al. 2018; Dai et al. 2014; Nobuta et al. 2012; Vestergaard \& Osmer 2009). In our fiducial framework, the average Eddington ratio slowly increases from values $\lambda\approx 0.1$ at $z\approx 1$ to values $\lambda\sim 0.6$ at $z\gtrsim 4$. The Eddington ratio distribution is quite broad, with a $1\sigma$ dispersion of $0.4$ dex almost independent of the redshift. The outcome from our approach is in good agreement with the observational estimates, although the latter, being mainly based on single-epoch estimators, are still subject to considerable uncertainties, especially toward high redshift. Note that recently in the literature a lot of attention has been paid to observational estimates of the Eddington ratio distribution as a function of host galactic properties, most noticeably stellar mass and specific SFR (e.g., Bongiorno et al. 2016; Georgakakis et al. 2017; Aird et al. 2018, 2022; Yang et al. 2019; Ananna et al. 2022; Carraro et al. 2020, 2022); however, the estimates are still subject to considerable uncertainties, especially at $z\gtrsim 0.5$ and for massive galaxies. The comparison with, and the interpretation of such distributions is beyond the scope of the present paper, and we defer it to a future work. 

In the above Figs. \ref{fig|agnlf} and \ref{fig|erdf} we also illustrate (dashed lines) the expected luminosity functions and average Eddington ratio when the dynamical friction mechanism is switched off and light BH seeds $\approx 10^2\, M_\odot$ are assumed (a value taken as representative for the most massive seeds of stellar origin). The results on the luminosity functions are almost indistinguishable from our fiducial case, since by construction our approach imposes that, with or without dynamical friction, the final BH masses must adhere to the same Magorrian relationship and are obtained within the same timescales set by the main sequence; in turn, this implies that the peak AGN luminosities are very close to each other. However, without dynamical friction, this is at the cost of increasing somewhat the average Eddington ratio, because the growth starts from a lighter seed.  Albeit in a statistical sense these higher values of $\lambda$ are still within the large dispersion of the observational data, the problem may be exacerbated for the very massive BHs $M_\bullet\gtrsim 10^9\, M_\odot$, and especially so at high $z\gtrsim 6$ whose formation would require $\lambda\sim$ a few (see Fig. \ref{fig|timevo} and related discussion above).

\begin{figure}
\centering
\includegraphics[width=\textwidth]{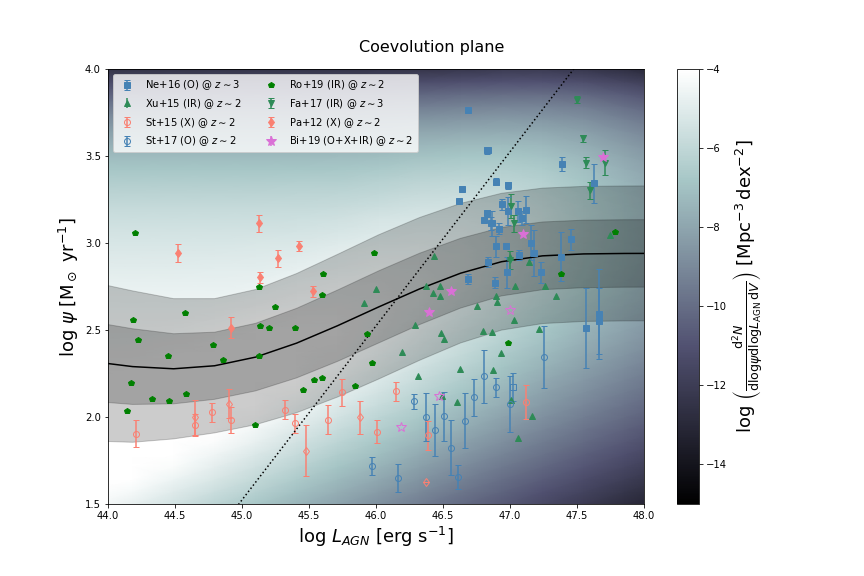}
\caption{The coevolution plane, namely the relationship between SFR of the host galaxy and the bolometric AGN luminosity $L_{\rm AGN}$ at a reference redshift $z\approx 2$. The intensity of the black and white background illustrates the number density of galaxies expected in the different portions of the diagram, and the solid line is the average relationship from our approach (dark and light grey shades represents the $1\sigma$ and $2\sigma$ dispersion). For reference, the dotted line represents the locus where the bolometric luminosity from the AGN and from the star formation in the host are equal. Data are from Netzer et al. (2015; squares), Xu et al. (2015; triangles), Stanley et al. (2015, 2017; circles), Fan et al. (2016; inverted triangles), Page et al. (2012; diamonds), Bianchini et al. (2019; stars), Rodighiero et al. (2019; pentagons). Symbol colors refer to observational selection in the optical (blue), X-ray (orange), IR (green), or mixed (magenta); moreover, filled symbols refer to detections, while empty symbols refer to stacking estimates.}\label{fig|coevplane}
\end{figure}

In Fig. \ref{fig|coevplane} we illustrate the coevolution plane at a reference redshift $z\approx 2$; this represents the number density of objects in the SFR $\psi$ vs. AGN bolometric luminosity $L_{\rm AGN}$ diagram (grey-scale color-coded); the average relationship and its $1-2\sigma$ scatter (solid line and shaded areas) are computed from such a distribution, taking into account the typical SFR detection threshold of present observations, around $\psi\approx 150\, M_\odot$ yr$^{-1}$. The distribution of objects in the coevolution plane is again determined mainly by the number density of galaxies with a given value of the SFR, implying that galaxies with higher SFRs are rarer, and by the time a galaxy spends in different AGN luminosity bins. The average SFR and its scatter, computed taking into account the typical SFR detection threshold mentioned above, stays roughly constant with AGN luminosity out to $L_{\rm AGN}\approx 10^{46}$ erg s$^{-1}$ and then slowly increases. Such a rise occurs just because, statistically, to achieve a higher AGN luminosity, the BH must reside in a more massive galaxy with a higher initial SFR. For comparison, in Fig. \ref{fig|coevplane} we report various observational determinations (see Page et al. 2012; Netzer et al. 2015; Stanley et al. 2015, 2017; Fan et al. 2016; Bianchini et al. 2019; Rodighiero et al. 2019) concerning different primary AGN selections in the optical, X-ray, IR or mixed (color-coded); detections are highlighted with full symbols and stacked data with open symbols. Our findings are remarkably consistent with observations, with the detections being distributed around the average relationship within its scatter, and the stacked measurements settling at the margin of the expected $2\sigma$ dispersion. 

\begin{figure}
\centering
\includegraphics[width=\textwidth]{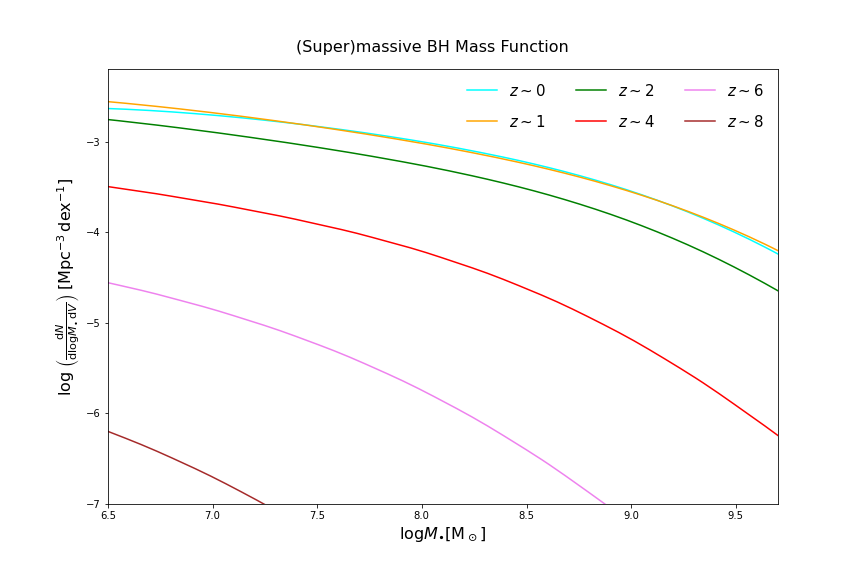}
\caption{The relic (super)massive BH mass function (see Eq. \ref{eq|BHMF}; solid lines) at different redshifts $z\approx 0$ (cyan), $1$ (orange), $2$ (green), $4$ (red), $6$ (violet), $8$ (brown).}\label{fig|BHMF_mf}
\end{figure}

In Fig. \ref{fig|BHMF_mf} we illustrate the (super)massive relic BH mass function as derived from the continuity equation Eq. (\ref{eq|BHMF}) at different redshifts $z\approx 0-8$ (color-coded). The redshift evolution is quite strong down to $z\approx 2$, with the knee (characteristic BH mass) strongly increasing from $M_\bullet\lesssim 10^7\, M_\odot$ at $z\gtrsim 8$ up to $M_\bullet\gtrsim 10^9\, M_\odot$ for $z\lesssim 2$; the evolution slows down considerably, especially at the high mass end, for $z\lesssim 2$, such that essentially below $z\approx 1$ the mass function undergoes only a minor evolution. 
 
\begin{figure}
\centering
\includegraphics[width=\textwidth]{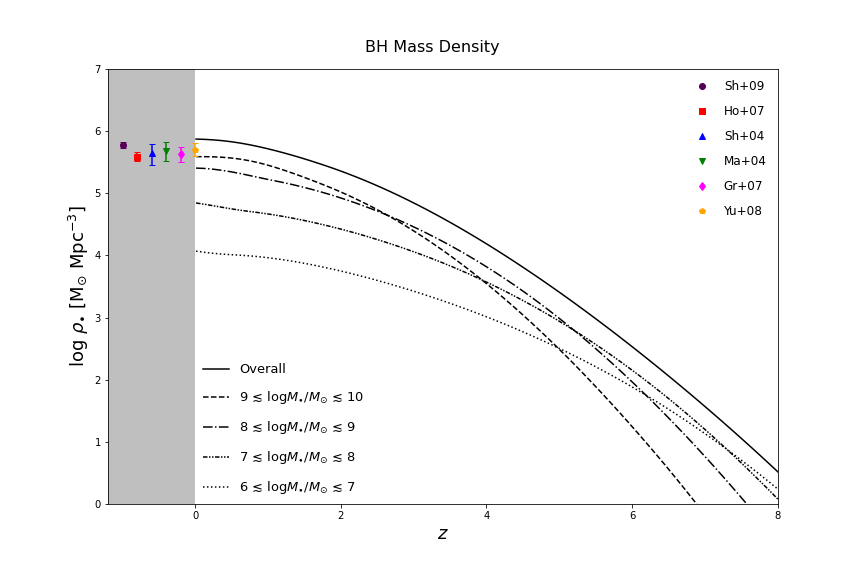}
\caption{The (super)massive relic BH mass density (see Eq. \ref{eq|BHdensity}) as a function of redshift $z$. The overall mass density is illustrated (solid) together with the contributions from the mass ranges $10^{6}\, M_\odot\lesssim M_\bullet\lesssim 10^{7}\, M_\odot$ (dotted), $10^{7}\, M_\odot\lesssim M_\bullet\lesssim 10^{8}\, M_\odot$ (triple-dot-dashed) and $10^{8}\, M_\odot\lesssim M_\bullet\lesssim 10^{9}\, M_\odot$ (dot-dashed) and $10^{9}\, M_\odot\lesssim M_\bullet\lesssim 10^{10}\, M_\odot$ (dashed). Observational estimates at $z\approx 0$ are from Shankar et al. (2009; brown circle), Hopkins et al. (2007; red square), Shankar et al. (2004; blue triangle), Marconi et al. (2004; green inverted triangle), Graham et al. (2007; magenta diamond), Yu \& Lu (2008; orange pentagon).}\label{fig|BHMF_bhmd}
\end{figure}

In Fig. \ref{fig|BHMF_bhmd} we show the related BH mass density computed after Eq. (\ref{eq|BHdensity}). It increases quite steeply from $\rho_\bullet\lesssim 10^{3}\, M_\odot$ Mpc$^{-3}$ at $z\gtrsim 6$ up to some $\rho_\bullet\gtrsim 10^{5}\, M_\odot$ Mpc$^{-3}$ at $z\lesssim 1$. The local BH mass density amounts to $\rho_\bullet\approx 6\times 10^5\, M_\odot$ Mpc$^{-3}$, in sound agreement with the available observational determinations (see Shankar et al. 2004, 2009: Hopkins et al. 2007; Marconi et al. 2004; Graham et al. 2007; Yu \& Lu 2008). Fig. \ref{fig|BHMF_bhmd} also displays the contribution to the mass density from different BH mass ranges, to highlight that at $z\lesssim 6$ and for $M_\bullet\lesssim 10^9\, M_\odot$, more massive BHs tend to accumulate their mass faster, displaying a kind of downsizing behavior.  

\begin{figure}
\centering
\includegraphics[width=\textwidth]{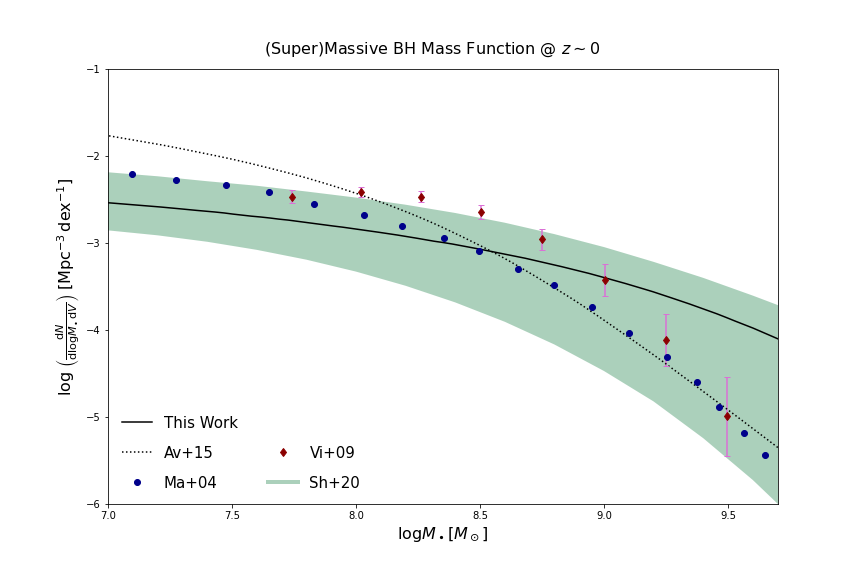}
\caption{The (super)massive BH mass function at $z\approx 0$. Solid line illustrates the outcome of our framework, while dotted line is the mass function originated when coupling the observed AGN luminosity functions with the average Eddington ratio adopted by Aversa et al. (2015). Observational estimates are from Marconi et al. (2004; blue circles), Vika et al. (2009; red diamonds), and Shankar et al. (2009, 2016, 2020; green shaded area); the latter reflects the overall uncertainty region when determining the BH mass function from the local stellar mass/velocity dispersion functions combined with various relationships of these observables with the BH mass.}\label{fig|BHMF_localmf}
\end{figure}

In Fig. \ref{fig|BHMF_localmf} we present the local BH mass function, and compare it with theoretical and observational estimates. In particular, the green shaded area refers to the uncertainty region in the current estimates of the BH mass function (see Shankar et al. 2016, 2020a), obtained when combining the local stellar mass/velocity dispersion functions with various literature relationships linking BH mass to stellar mass/velocity dispersion of the host. We also report for comparison the classic estimates by Marconi et al. (2004; see also Shankar et al. 2009) via a simplified continuity equation approach, and by Vika et al. (2009) via an object-by-object analysis of the BH mass-host luminosity relationship. Our mass function is in agreement with most determinations for BH masses $M_\bullet\lesssim$ some $10^8\, M_\odot$. At the high-mass end it lies well within the Shankar et al. (2020a) uncertainty region, but it declines substantially slower with respect to the classic estimates by Marconi et al. (2004) and Vika et al. (2009). 

We stress that to obtain a BH mass function with a steep behavior at the high mass end is a non-trivial task. Specifically, in our framework we determine $\lambda$ from the empirical Magorrian relation and main sequence timescale, obtaining values $\lambda<1$ that are in good agreement with the observed Eddington ratios; we also predict AGN luminosity functions closely matching the data. However, when inserted into the continuity equation these low $\lambda$ values originate a rather flat BH mass function at the high mass end since large BH masses correspond to moderate peak AGN luminosities (approximately $L_{\rm AGN}\propto \lambda\, M_{\bullet}$ holds) falling in the rather flat portion of the luminosity function. Even the slightly higher $\lambda$ values we obtain when switching off dynamical friction (see dashed line in Fig. \ref{fig|erdf}) are not sufficient to appreciably steepen the BH mass function, which features a high mass end similar to our fiducial case. Contrariwise, in other literature studies (e.g., Aversa et al. 2015; Shen et al. 2020) a steep behavior of the mass function is enforced by starting from the observed AGN luminosity functions (not self-consistently predicting them, as in this work) and by assuming values $\lambda\gtrsim 1$ designed on purpose. For example, the redshift-dependent parameterization $\lambda(z)\approx 4\,\{1-0.5\times{\rm erfc}[(z-2)/3]\}$ proposed by Aversa et al. (2015) works quite well in producing a steep BH mass function, but at the price of assuming $\lambda$ values somewhat in tension with the observed average Eddington ratios (see dotted lines in Figs. \ref{fig|erdf} and \ref{fig|BHMF_localmf}). Insisting on such high $\lambda$ values in a self-consistent approach while maintaining a good prediction of the AGN luminosity functions is still possible, but requires BH growth timescales $\lesssim 100$ Myr, much shorter than derived via the main sequence prescription.

\subsection{Robustness of results against main assumptions}\label{sec|comparison}

\begin{figure}
\centering
\includegraphics[width=\textwidth]{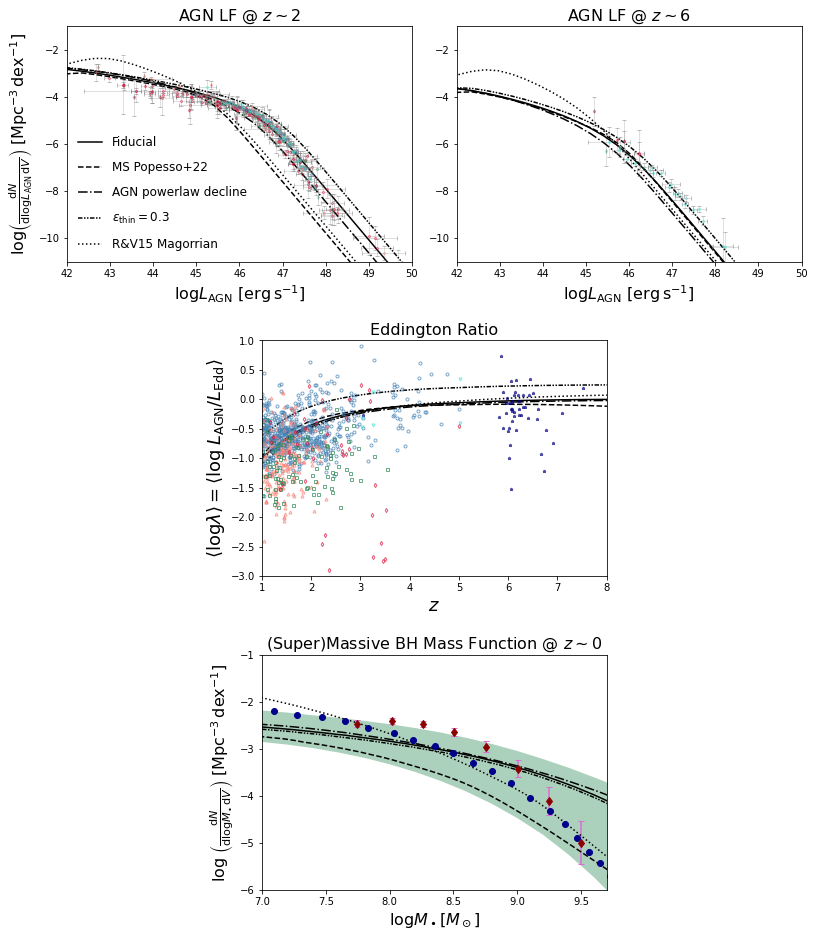}
\caption{Dependence of our results concerning AGN luminosity function at $z\sim 2$ (top left) and $z\sim 6$ (top right), average Eddington ratio as a function of redshift (middle), and local supermassive BH mass function (bottom) to various assumptions/relationships employed in this work. In all panels solid lines refer to our fiducial assumptions, dashed lines to our results when the main sequence relation by Popesso et al. (2022) is used in place of Speagle et al. (2014), dot-dashed line to our results when the BH accretion rate curves is characterized by a powerlaw decline instead of an exponential one, dotted lines to our results when the Magorrian relation by Reines \& Volonteri (2015) for AGNs is employed in place of the one by Shankar et al. (2016, 2020a), and triple-dot-dashed line to our results when the thin disk efficiency $\epsilon_{\rm thin}\approx 0.3$ is adopted instead of our fiducial value $\epsilon_{\rm thin}\approx 0.15$. See Sect. \ref{sec|comparison} for details.}\label{fig|BHMF_comparison}
\end{figure}

In Fig. \ref{fig|BHMF_comparison} we highlight the dependence of our results concerning the AGN luminosity function, redshift evolution of the average Eddington ratio, and local (super)massive BH mass function on various assumptions/relationships used in our reference framework. 

First, we vary the main sequence relationship, switching from Speagle et al. (2014) to the recent determination by Popesso et al. (2022). In analogy with Eq. (\ref{eq|MS_sp14}), this can be rendered as
\begin{equation}\label{eq|MS_Po22}
\log \frac{\psi_{\rm MS}(M_\star,z)}{M_\odot\, {\rm yr}^{-1}} \approx (-27.58+0.26\, t_z)+(4.95-0.04\,t_z)\, \log \frac{M_\star}{M_\odot}-0.2\,\log^2 \frac{M_\star}{M_\odot}\;.
\end{equation}
With respect to the almost linear relation by Speagle et al. (2014), the above is characterized by a steepening toward the lower stellar masses and a progressive flattening toward higher stellar masses, that have some relevance in galaxy formation since they may be interpreted as the effects of stellar feedback and mass quenching, respectively (e.g., Lapi et al. 2018; Daddi et al. 2022).

Second, we vary the shape of the declining portion of the accretion rate curve in Eq. (\ref{eq|BHaccrate}). In particular, we switch from an exponential to a scale-free, powerlaw shape
$M_{\bullet,\rm acc}(\tau)=\dot M_{\bullet,\rm acc}(\tau_{\rm b})\, \left(\tau/\tau_{\rm b}\right)^{-\omega}$ for $\tau>\tau_{\rm b}$. Here $\omega>1$ rules the steepness of the decline, and we set $\omega\approx 2.5$ as in Shen (2009). Correspondingly, the overall BH growth at late times (cf. Eq. \ref{eq|BHrate}) follows
\begin{equation}
M_{\bullet}(\tau|\psi,z)= M_{\bullet}(\tau_{\rm b})\, \left[1+\cfrac{\tau_{\rm b}}{\tau_{\rm ef}}\, \cfrac{1-(\tau/\tau_{\rm b})^{1-\omega}}{\omega-1}\right]\;\;\;\;, \;\;\;\;\;\;\;\;\;\;\;\;\tau>\tau_{\rm b}
\end{equation}
and the relic mass for $\tau\gg\tau_{\rm b}$ reads $M_{\bullet,\rm relic}(\psi,z)=M_{\bullet}\,[1+\tau_{\rm b}/(\omega-1)\,\tau_{\rm ef}]$. Such a powerlaw behavior is often adopted in empirical BH evolution models and generically ascribed to a residual accretion related to viscosity in a thin accretion disk (e.g., Yu \& Lu 2008; Shen 2009); it has also been claimed to be consistent with a few numerical simulations present in the literature (see discussion by Habouzit et al. 2022). 

Third, we vary the adopted Magorrian relationship (cf. Eq. \ref{eq|MBHMbulge}) from the debiased determination by Shankar et al. (2016, 2020a) based on dynamical BH masses to that by Reines \& Volonteri (2015) based on single-epoch virial estimators for locally active BHs (calibrated on a subsample of reverberation-mapped AGNs): 
\begin{equation}
\log \frac{M_{\bullet,{\rm Mag}}}{M_\odot}(M_\star,z) \approx 7.45+1.05\,\log\frac{M_\star}{10^{11}\, M_\odot}+\eta\,\log(1+z)\;,
\end{equation}
where for consistency we retain the same redshift dependence adopted in Eq. (\ref{eq|MBHMbulge}). 

Finally, we vary the radiative efficiency $\epsilon_{\rm thin}$ of the thin disk regime (see Eq. \ref{eq|radeff_av15}) from our fiducial value $0.15$ to $0.3$; the latter is close to the limit applying for maximally spinning BHs. In fact, some theoretical works (e.g., Volonteri et al. 2013; Sesana et al. 2014; Griffin et al. 2019; Izquierdo-Villalba et al. 2020) have pointed out that the population of high-$z$ BHs might be maximally spinning, so it is interesting to check the effect of this assumption especially on the AGN luminosity function at high redshift $z\gtrsim 6$.

Fig. \ref{fig|BHMF_comparison} shows that the most critical assumptions are, not surprisingly, the adopted main sequence and Magorrian relationships, that clearly affect the timescale of BH growth and the final BH masses, hence the resulting AGN luminosity function and BH mass function. As for the Popesso main sequence, it causes both a reduced number density of galaxies with high SFR, and a smaller stellar mass at a given SFR. This yields smaller BH masses hence a lower and steeper BH mass function. At the same time, with the Popesso main sequence shorter timescales are available for BH growth, implying minor variations in the Eddington ratio and correspondingly lower luminosities. As for the Magorrian relation by Reines \& Volonteri (2015), it is flatter than our reference case and tends to yield lower BH masses for stellar masses $M_\star\gtrsim$ a few $10^{10}\, M_\odot$, and viceversa. Overall, this naturally originates an AGN luminosity function and a local BH mass function pumped at the faint end and depressed at the bright one, while the change in the average Eddington ratio is minor. Adopting a power-law shape of the declining portion in the BH accretion rate curve affects somewhat the AGN luminosity functions, while the impact on the Eddington ratio distribution and on the BH mass function is limited. Finally, we also highlight that adopting a high value $\epsilon_{\rm thin}\approx 0.3$ of the thin disk radiative efficiency implies higher Eddington ratio $\lambda$. This is seen by combining Eqs. (\ref{eq|tauef}) and (\ref{eq|radeff_av15}) given that $\tau_{\rm ef}$ stays put since it is determined for a final BH mass by the Magorrian relation and the main sequence timescale. In the end this originates higher AGN luminosity functions, which better agree with observational estimates for $z\gtrsim 6$; this is particularly interesting since, as mentioned above, higher efficiencies associated to quickly spinning BHs are mostly expected toward such high redshifts.

\subsection{The Overall BH mass function}

\begin{figure}
\centering
\includegraphics[width=\textwidth]{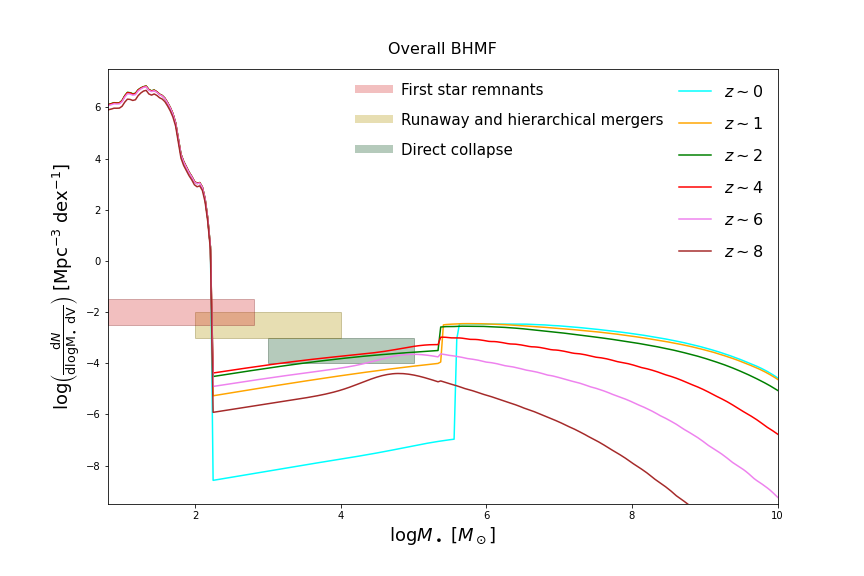}
\caption{The overall BH mass function from our semi-empirical framework, from the stellar to the intermediate to the (super)massive regime, at different redshifts $z\approx 0$ (cyan), $1$ (orange), $2$ (green), $4$ (red), $6$ (purple), and $8$ (brown). The colored boxes illustrate the mass and density ranges from other seed formation channels (see Volonteri et al. 2021): remnants of the first massive pop-III stars (red box), direct collapse of primordial gas clouds (green box), and runaway stellar or BH mergers in compact primeval star clusters (yellow box).}\label{fig|BHMF_overall}
\end{figure}

In Fig. \ref{fig|BHMF_overall} we illustrate the overall black hole mass function, from the stellar to the (super)massive regime, over more than ten orders of magnitude in BH mass. The stellar regime for $M_\bullet\lesssim 10^2\, M_\odot$ is taken from paper I and strictly associated to the star formation process in galaxies. In our framework, the intermediate mass regime $M_\bullet\sim 10^{2-5}$ is mainly associated to the formation of heavy BH seeds by migration of stellar BHs via gaseous dynamical friction at the center of star-forming galaxies; the migrating stellar mass BHs are a very tiny fraction of the total number, so that the number density of these intermediate BHs is substantially lower than the stellar one. Finally, the (super)massive regime $M_\bullet\sim 10^{6-10}\, M_\odot$ is mainly populated by the BHs that have grown to large masses (from heavy seeds) via Eddington-type gas disk accretion. Most of such massive BHs are \emph{active} at high redshifts $z\gtrsim 6$, so that the BH mass function in the intermediate and (super)massive regime is continuously connected. On the other hand, moving toward lower redshifts the mass function in the (super)massive range increases because \emph{relic} BHs grown by disk accretion accumulate, while the number of intermediate mass BHs diminishes since the dynamical friction process becomes less efficient and the overall production of stellar mass BHs also lowers (following the progressive decline in the amount of star-formation within galaxies). This is at the origin of the discontinuity (or gap; see Trinca et al. 2022 and Spinoso et al. 2022 for a similar behavior) between the intermediate and (super)massive mass function around $M_\bullet\lesssim 10^6\, M_\odot$; it is pleasing that this transition occurs at around the typical value usually considered to separate intermediate from supermassive BHs.

For reference, in Fig. \ref{fig|BHMF_overall} we have also illustrated as
colored boxes the mass and density ranges expected from other classic seed formation channels (taken from Volonteri et al. 2021, see their Fig. 1): remnants of the first massive pop-III stars (red box), direct collapse of primordial gas clouds (green
box), and runaway stellar or BH mergers in compact primeval star clusters (yellow box). 
These distributions mainly originates in (proto)galaxies at $z\gtrsim 10$, and are then progressively eroded (but not substantially refurnished) at lower redshifts, when the seeds merge together or accrete gas and become more massive BHs (e.g., Mayer \& Bonoli 2019; Volonteri et al. 2021; Trinca et al. 2022; Spinoso et al. 2022). This is at variance with our framework, where heavy seeds are continuously produced across cosmic times by the migration and merging of stellar-mass BHs associated to star formation in galaxies. In view of the above, if present, such classic seed formation channels are expected to enhance somewhat the BH mass function in the range $M_\bullet\sim 10^{2-5}\, M_\odot$ especially at redshifts $z\gtrsim 8$. At later cosmic times, classic formation channels will feature a substantially eroded distribution in the intermediate mass range, so that their impact on our BH mass function should be minor. However, in a future work it would be interesting to perform a detailed investigation of the cooperative action of all these seed formation mechanisms across cosmic history.

\section{Summary and Outlook}\label{sec|summary}

In this work we have provided an ab-initio computation of the (super)massive BH mass function across cosmic times (see Fig. \ref{fig|Schematic}). To this purpose, we have started from the redshift-dependent galaxy statistics (constituted by the SFR functions) and have modeled the joint evolution of the central BH mass and the stellar mass of the host (see Fig. \ref{fig|timevo}). We have considered two mechanisms to grow the central BH, that are expected to cooperate in the high-redshift star-forming progenitors of local massive galaxies. One is the gaseous dynamical friction envisaged by Boco et al. (2020), that can cause the migration of stellar-mass BHs originated during the intense bursts of star formation toward the gas-rich central regions of the host progenitor galaxies; this leads to the buildup of an heavy BH seeds $\lesssim 10^5\, M_\odot$ within short timescales $\lesssim$ a few $10^7$ yr. The second mechanism is the standard Eddington-type gas disk accretion onto the heavy seed, through which the central BH can become (super)massive within the typical star-formation timescales $\lesssim 1$ Gyr of the host galaxy, as set by the galaxy main sequence. We have self-consistently combined these mechanisms to compute the overall growth rate functions of the central (super)massive BHs (see Fig. \ref{fig|BHMF_grf}).

We have validated our approach by consistently reproducing the observed redshift-dependent bolometric AGN luminosity functions (Fig. \ref{fig|agnlf}), the observed Eddington ratio distributions (Fig. \ref{fig|erdf}), and the observed relationship between the star-formation of the host galaxy and the bolometric luminosity of the accreting central BH (Fig. \ref{fig|coevplane}). We have then derived the relic (super)massive BH mass function (Fig. \ref{fig|BHMF_mf}) and BH mass density (Fig. \ref{fig|BHMF_bhmd}) via a generalized continuity equation approach, finding a pleasing agreement with the most recent observational estimates at $z\approx 0$ (Fig. \ref{fig|BHMF_localmf}). All in all, we have found that the present (super)massive BH mass density amounts to $\rho_\bullet\approx 6\times 10^5\, M_\odot$ Mpc$^{-3}$, in accord with available estimates. 

We have stressed that in the absence of the dynamical friction process, statistical observables like the AGN luminosity functions are not substantially affected, since most of the BH mass is accumulated in the gas disk accretion phase. However, to attain BH masses $\gtrsim 10^9\, M_\odot$ within the typical star-formation duration $\lesssim 1$ Gyr of the host galaxy without such dynamical friction process is challenging,  especially at high redshifts $z\gtrsim 6$ or for overmassive BHs that are upper outliers of the average Magorrian relationship. In such a case, the BH growth must proceed at appreciably high Eddington ratios $\lambda\gtrsim 1$ and/or starting from heavy BH seeds $M_\bullet\sim 10^{3-5}\, M_\odot$. The first instance can be partially justified theoretically but struggles somewhat against present observational estimates, the second would require a specific mechanism, alternative to gaseous dynamical friction, designed to obtain such massive seeds. 

Finally, putting together the results from paper I and the present work, we have reconstructed the overall BH mass function from the stellar to the (super)massive regime over more than ten orders of magnitude in BH mass.  At the same time, we have provided a robust theoretical basis for a physically-motivated heavy seed distribution as a function of redshift. At variance with classic seed production channels, in our framework the heavy seed distribution is time-dependent: heavy seeds are continuously produced by the merging of light seeds originated from star formation via the gaseous dynamical friction mechanism; but they also grow via standard Eddington-type accretion, and soon leave the intermediate mass regime to become (super)massive. It would be extremely interesting to implement such a time-dependent seed distribution in analytic and numerical models of BH formation and evolution. 

In a future perspective, our semi-empirical approach could be exploited to populate a $N-$body simulation, in order to build up a mock catalog encapsulating the three-dimensional spatial distribution and clustering of heavy seeds and of (super)massive BHs within their galactic hosts (e.g., Allevato et al. 2021). Another development could be a more detailed comparison of the properties of (super)massive BHs and host star forming galaxies, for example in terms of Eddington distributions as a function of BH environment and host galaxy properties (SFR, stellar mass, nuclear obscuration, etc.; e.g., Aird et al. 2018; Ananna et al. 2022). 
Moreover, we plan to work out predictions for upcoming or future observations via space instruments like JWST and \textit{Athena}. Specifically, young BHs lying at the center or wandering in the nuclear regions of dusty starforming hosts may be detectable, even in the early stages of growth, via their X-ray and/or strongly extincted UV emissions; the latter could constitute a probe for the existence and abundance of intermediate mass BHs and could provide a characterization of their main growth mechanisms. Finally, we aim to exploit the BH mass function derived here to estimate the rate of (super)massive BH mergers. Although their effect on the overall mass function is expected to be mild and confined at the very massive end and late cosmic times, these events can constitute powerful sources of gravitational waves (e.g., Barausse \& Lapi 2021). Thus we will provide detailed forecasts for their detectability by the Laser Interferometer Space Antenna mission and by present and future Pulsar-Timing Array experiments.

\begin{appendix}

\section{Migration of stellar BH\lowercase{s} via gaseous dynamical friction}\label{sec|app_dynfric}

In this Appendix we recall some details of the mechanism envisaged by Boco et al. (2020) to grow heavy seeds via migration of stellar mass BHs due to gaseous dynamical friction.

In the local Universe, supermassive BHs are hosted at the center of massive spheroidal galaxies. Thus their heavy seeds must have formed in the progenitors of such systems at intermediate/high redshifts, which are known to be dusty star-forming galaxies. These objects, detected and investigated mainly in the far-IR/(sub)mm band by ground-based interferometers like ALMA, feature large SFRs $\psi\gtrsim 10^2-10^3$ M$_\odot$ yr$^{-1}$ and huge molecular gas reservoirs $M_{\rm gas}\gtrsim 10^{10}-10^{11}$ M$_\odot$ concentrated in a compact region of a few kpc (see Scoville et al. 2014, 2016; Ikarashi et al. 2015; Simpson et al. 2015; Barro et al. 2016; Spilker et al. 2016; Tadaki et al. 2017a, 2017b, 2018;  Lang et al. 2018; Talia et al. 2018, 2021; Smail et al. 2021). These conditions are prompt for the efficient sinking of many compact objects toward the nuclear regions via gaseous dynamical friction (e.g., Ostriker 1999; Sanchez-Salcedo \& Brandenburg 2001; Escala et al. 2004; Tanaka \& Haiman 2009; Tagawa et al. 2016; Boco et al. 2020).

Specifically, Boco et al. (2020) have run a series of dynamical simulations and derived a fitting formula for the corresponding migration timescale of stellar mass BHs:
\begin{equation}
\tau_{\rm DF}\approx \mathcal{N}\left(\frac{m_\bullet}{100\,\rm M_\odot}\right)^a \left(\frac{M_{\rm gas}}{10^{11}\,M_\odot}\right)^b \left(\frac{R_{\rm e}}{1\,\rm kpc}\right)^c \left(\frac{j}{j_c}\right)^\beta \left(\frac{r_c}{10\,\rm pc}\right)^\gamma\; ;
\label{eq|tauDF}
\end{equation}
here $M_{\rm gas}$ is the total gas mass, $R_e$ is the half mass radius of the gas distribution, $m_\bullet$ is the mass of the migrating compact object, $\epsilon$ and $j$ are the initial specific energy and angular momentum of the compact object, $r_c(\epsilon)$ is the circular radius that the compact object would have if it were on a circular orbit with energy $\epsilon$, and $j_c(\epsilon)$ is the angular momentum associated to that orbit. The precise values of the exponents $(a,b,c,\beta,\gamma)$ and of the normalization factor $\mathcal{N}$ depend on the specific shape of the gas density profile. In the present work we adopt the fiducial setup of Boco et al. (2020), namely, a 3D Sersic gas density profile $\rho(r)\propto r^{-\alpha}\, e^{-k\,(r/R_e)^{1/n}}$ with $n=1.5$, $\alpha=1-1.188/2n+0.22/4n^2\sim 0.6$ and half-mass radius $R_{\rm e}\sim 1$ kpc. Then the values for the parameters in Eq. (\ref{eq|tauDF}) read $a\approx-0.95$, $b\approx0.45$, $c\approx-1.2$, $\beta\approx 1.5$, $\gamma\approx 2.5$ and $\mathcal{N}\approx 3.4\times 10^8$ yr. The effect of different setups on the dynamical friction timescale is discussed in Boco et al. (2020).

Given the high SFR ongoing in the progenitors of local spheroidal galaxies, a lot of stars and compact remnants are formed in a short timescale within the nuclear regions. We assume that stars are initially distributed in space as the gas density profile $\rho$, so that the probability distribution for a star to be born at distance $r$ from the galactic center is ${\rm d}p/{\rm d}r\propto r^2\, \rho(r)$.
After $\lesssim 10^7\,\rm yr$ massive stars ($m_\star\gtrsim 7-8\,M_\odot$) undergo a supernova explosion possibly leaving a stellar mass BH. We assume that the latter follow the same velocity distribution of the progenitor stars, which is in turn related to that of the star-forming molecular gas cloud. In particular, we assume a Gaussian distributions of radial and tangential velocities:
${\rm d}p/{\rm d}v_{r,\theta}\propto e^{-v_{r,\theta}^2/2\sigma^2}$
with dispersion $\sigma(r)$ found by solving the isotropic Jeans equation:
$\sigma^2(r)\propto \rho^{-1}(r)\,\int_r^\infty\,{\rm d}r'\, \rho(r')\,r'^{-2}\, \int_0^{r'}{\rm d}r''\, r''^2\,\rho(r'')$.
From these distributions the initial positions and velocities of stellar BHs, their initial energy and angular momentum can be easily extracted.

For a galaxy with spatially-averaged SFR $\psi$, we compute the associated relic stellar mass $M_\star$ from the well-established galaxy main sequence relationships (e.g., Speagle et al. 2014) and then estimate the initial gas mass $M_{\rm gas}$, entering the dynamical friction timescale, via the redshift-dependent $M_{\rm gas}-M_\star$ relation from abundance matching techniques (see Moster et al. 2013, 2018; Aversa et al. 2015; Shi et al. 2017; Behroozi et al. 2019). Finally, the dynamical friction timescale $\tau(m_\bullet, r, v_R,v_\theta)$ can be 
computed from Eq. (\ref{eq|tauDF}), and the convolution of the stellar BH birthrate with the aforementioned distributions of initial position and velocity yields the migration rates according to Eq. (\ref{eq|dynfricrate}).

\section{Continuity equation}\label{sec|app_conteq}

In this Appendix we provide some details on how to solve the continuity equation in the integral formulation, along the lines envisaged by Yu \& Lu (2004, 2008) and Aversa et al. (2015). The continuity equation links the BH mass function $N(M_\bullet,t)\equiv {\rm d}N/{\rm d}M_\bullet\,{\rm d}V$ and the BH growth rate function $N(\dot M_\bullet,t)\equiv {\rm d}N/{\rm d}\dot M_\bullet\,{\rm d}V$ according to
\begin{equation}\label{eq|intcont}
N(\dot M_\bullet,t)=\int_0^\infty{\rm d}M_\bullet\, [\partial_t N(M_\bullet,t)-S(M_\bullet,t)]\, \sum_i\cfrac{{\rm d}\tau_i}{{\rm d}\dot M_\bullet}(\dot M_\bullet|M_\bullet,t)\, \Theta_{\rm H}[\dot M_\bullet\leq \dot M_{\bullet}(M_\bullet,t)]\; ;
\end{equation}
here $t$ is the cosmic time, $\tau$ the proper time since the triggering of the BH growth, ${\rm d}\tau/{\rm d}\dot M_\bullet$ is the time spent by the BH with final mass $M_\bullet$ in a bin of growth rate ${\rm d}\dot M_\bullet$ given a growth curve $\dot M_\bullet(\tau|M_\bullet, t)$; the summation allows for multiple solutions $\tau_i$ of the equation $\dot M_\bullet(\tau|M_\bullet, t)=\dot M_\bullet$. In addition, $S(M_\bullet,t)$ is a source term accounting for (super)massive BH mergers, that we neglect hereafter. Finally, the Heaviside step function $\Theta_{\rm H}[\cdot]$ specifies that the growth rate must be smaller than the maximum value $\dot M_{\bullet}(M_\bullet, t)$ for a given final BH mass.

Multiplying both sides of Eq. (\ref{eq|intcont}) by $\dot M_\bullet$ and writing explicitly the constraints implied by the step function yields
\begin{equation}
N(\log \dot M_\bullet,t)=\int_{M_\bullet(\dot M_{\bullet},t)}^\infty\, {\rm d}M_\bullet\, \partial_t N(M_\bullet,t)\, \sum_i\cfrac{{\rm d}\tau_i}{{\rm d}\log \dot M_\bullet}\; ,
\end{equation}
where $M_\bullet(\dot M_\bullet,t)$ is the minimum BH mass that has accreted at $\dot M_\bullet$. After differentiating both sides by $\log \dot M_\bullet$ one obtains
\begin{equation}
\partial_{\log \dot M_\bullet}N(\log \dot M_\bullet,t)= - \partial_t N(\log M_\bullet,t)\, \left|\partial_{\log \dot M_\bullet}\log M_\bullet\right|\, \sum_i\cfrac{{\rm d}\tau_i}{{\rm d}\log \dot M_\bullet}\; .
\end{equation}
Rearranging the expression and integrating over cosmic time, one finally can write the solution of the continuity equation in the form
\begin{equation}
N(\log M_\bullet,t) = -\int_0^t\,\cfrac{{\rm d}t'}{\sum_i{\rm d}\tau_i/{\rm d}\log \dot M_\bullet}\; \left.\cfrac{\partial_{\log \dot M_\bullet}N(\log \dot M_\bullet,t')}{\partial_{\log \dot M_\bullet}\log M_\bullet}\right|_{\dot M_\bullet=\dot M_\bullet(M_\bullet,t')}\; ,
\end{equation}
where all the integrand is calculated at the maximum growth rate $\dot M_\bullet(M_\bullet,t)$ for a given BH mass. This is a very general solution of the continuity equation, that holds even when the parameters defining the growth curve, e.g. the $e-$folding time, depend on $M_\bullet$, $\dot M_\bullet$ and cosmic time $t$.

As a simple application, consider the very special case when the growth of the BH occurs by gas accretion in an Eddington-limited regime (with constant $e-$folding time $\tau_{\rm ef}$ independent of accretion rate and cosmic time) up to a time $\tau_{\rm b}$, so that $\dot M_\bullet(\tau) = M_\bullet(\tau)/\tau_{\rm ef}$ and $M_\bullet(\tau) = M_\bullet (\tau_{\rm b})\, e^{(\tau-\tau_{\rm b})/\tau_{\rm ef}}$ for $\tau\leq \tau_{\rm b}$ while $\dot M_\bullet = 0$ and $M_\bullet (\tau)=M_\bullet(\tau_{\rm b})$ for $\tau>\tau_{\rm b}$. In such an instance one has that the maximum BH accretion rate attained by a BH with final mass $M_\bullet$ is $\dot M_\bullet(M_\bullet) = M_\bullet/\tau_{\rm ef}$, hence $\left|\partial_{\log \dot M_\bullet}\log M_\bullet\right|=1$. Moreover, $\sum_i{\rm d}\tau_i/{\rm d}\log \dot M_\bullet = \tau_{\rm ef}\, \ln(10)$. All in all, the solution writes
\begin{equation}
N(\log M_\bullet,t) = -\int_0^t\,\cfrac{{\rm d}t'}{\tau_{\rm ef}\, \ln(10)}\; \left.\partial_{\log \dot M_\bullet}N(\log \dot M_\bullet,t')\right|_{\dot M_\bullet=M_\bullet/\tau_{\rm ef}}\; ,
\end{equation}
which is the classic expression derived by Marconi et al. (2004) and Shankar et al. (2004).

\end{appendix}

\begin{acknowledgements}
We thank the referee for a competent and constructive report. We acknowledge S. Bonoli, D. Donevski, G. Rodighiero and T. Ronconi for interesting discussions. This work has been supported by the EU H2020-MSCA-ITN-2019 Project 860744 `BiD4BESt: Big Data applications for black hole Evolution STudies'. AL, AB and MM acknowledge funding from the PRIN MIUR 2017 prot. 20173ML3WW, `Opening the ALMA window on the cosmic evolution of gas, stars and supermassive black holes'.
\end{acknowledgements}

\newpage

\newpage

\newpage

\newpage

\newpage

\newpage

\newpage

\newpage

\newpage

\newpage

\newpage

\end{document}